\documentclass[aps,prd,superscriptaddress,nofootinbib,showpacs,10pt]{revtex4}

\pdfoutput=1
\usepackage{graphicx}
\usepackage{hyperref}
\usepackage{ulem}
\usepackage{color}
\usepackage{slashed}
\usepackage{amsmath}
\usepackage{amssymb}
\definecolor{My_red}        {cmyk}{0.00,1.00,1.00,0.20}

\usepackage{multirow}
\usepackage{makecell}

\def\ra{\rightarrow}

\def\ra{\rightarrow}

\def\f{\frac}

\def\bwt{\begin{widetext}}
\def\ewt{\end{widetext}}
\def\be{\begin{equation}}
\def\ee{\end{equation}}
\def\bary{\begin{array}}
\def\eary{\end{array}}
\def\bit{\begin{itemize}}
\def\eit{\end{itemize}}

\def\ra{\rightarrow}

\def\su5u1{SU(5) \times U(1)}
\def\fsu5u1{SU(5) \times U(1)'}
\def\so10{SO(10)}
\def\sq20{SO(10) \times SO(10)}

\usepackage{ulem,fancyvrb} 
\usepackage[dvipsnames]{xcolor}

\begin{document}

\title{Missing particle associated with two bottom quarks at the LHC: 
\\
Mono-$b$ versus 2$b$ with razor variables}

\author{Ning Chen}
\email[E-mail: ]{chenning@ustc.edu.cn}
\affiliation{Department of Modern Physics, \\ University of Science and Technology of China, Hefei, Anhui 230026, China}
\affiliation{Kavli Institute for Theoretical Physics China (KITPC),
Institute of Theoretical Physics, Chinese Academy of Sciences,
Beijing 100190, P. R. China}

\author{Zhaofeng Kang}
\email[E-mail: ]{zhaofengkang@gmail.com}
\affiliation{School of Physics, Korea Institute for Advanced Study,
Seoul 130-722, Korea}
\affiliation{Kavli Institute for Theoretical Physics China (KITPC),
Institute of Theoretical Physics, Chinese Academy of Sciences,
Beijing 100190, P. R. China}

\author{Jinmian Li}
\email[E-mail: ]{phyljm@gmail.com}
\affiliation{School of Physics, Korea Institute for Advanced Study,
Seoul 130-722, Korea}
\affiliation{ARC Centre of Excellence for Particle Physics at the Terascale, Department of Physics, University of Adelaide, Adelaide, SA 5005, Australia}

\date{\today}

\begin{abstract}

The extended Higgs sector, such as by a second Higgs doublet of type-II, provides portals to dark sector which contains missing particles at the LHC, e.g., dark matter (DM) particles. 
In this paper, working in the simplified model and taking into consideration the wide decay width effect of the mediator, we analyze the characteristic signatures of mono-$b$+MET and 2$b$+MET at the LHC. 
The latter signature was believed to be ineffective. 
While we found that, with the aid of razor shape analysis, it should be as important as the mono-$b$ signature. 
In the region of relatively low mediator mass (below a few hundred GeV), by requiring the signal to background ratio greater than a few percent, the 2$b$-tagged razor analysis has comparable sensitivity to the mono-$b$ search; it is even better for mediator lighter than $\sim 200$ GeV.

\end{abstract}

\maketitle

\section{Introduction}

Hunting for dark matter (DM) is one of the main object of the current and future LHC experiments. 
However, owing to the lack of knowledge of interactions between the DM and the Standard Model (SM) particles, there are a huge pool of models to ``explain" the existence of DM and a clear prediction of DM signature at the LHC is lost.~\footnote{This motivates the use of effective operator approach~\cite{Cao:2009uw,Goodman:2010ku}, grounded on the integrating out heavy mediators that connect DM and quarks. 
But it can only describe a subset of models and a lot of information may be lost~\cite{Buchmueller:2013dya}. 
Then including the mediator and working in the simplified model should be a better setup.} 
One popular conjecture on the DM interaction is via the SM Higgs portal. 
In the domain of new physics, the Higgs sector may not just refer to the SM Higgs doublet of $\Phi_2$. 
Some popular extensions include a second Higgs doublet, a SM singlet scalar or both, and so on. 
It is of interest to consider the Higgs portal in the context of a broad Higgs sector. 
DM particles behave as missing transverse energy in the collider searches, which do not differ from any other particles that are neutral and long-lived. 
Therefore, to widen the use of this study, we will not restrict to DM, while generically refer DM as missing particle. 
Such a methodology avoids the conventional DM constraints, so that we just focus on the collider analysis.

In this paper, we consider DM interacting with the visible sector via a two-Higgs-doublet (2HDM) like portal, or more exactly $\Phi_1$-like portal. 
$\Phi_1$ has interactions with the SM fermions as those in the type-II 2HDM~\cite{Branco:2011iw}. 
The meaning of ``like" will be explained in the text. 
A lot of papers have studied such a portal~\cite{2HDM:DM,2HDM:DM0} (in particular, a ``derivation" of such a model in the framework of scale invariance~\cite{Guo:2014bha}), whereas a specific LHC study is still absent. 
Therefore, exploring the DM signatures at the LHC based on the $\Phi_1$-like portal is well motivated and timely. 
For the corresponding searches, we can take the advantage of the $gg\to b\bar b \Phi_1(\to {\rm DM+ DM})$ process, which is enhanced by large $\tan\beta$ and  furnishes possibly two visible particles, i.e. a pair of bottom quarks in the final state. 
The resulting signatures of mono-$b$ jet plus large missing transverse momentum (MET) and 2$b$-jets plus MET are different from the usual mono-jet signature~\cite{monojet}, where the later was based on effective operators like $\bar q q \bar\chi \chi$, with $q$ being light quarks and $\chi$ being a fermionic DM field. 
This paper aims to analyze these signatures, which have not received much attention yet. 

The mono-$b$ analysis was initiated in Ref.~\cite{Lin:2013sca} based on the scalar operator ${\cal O}_b=\bar bb\bar\chi \chi$ and followed by Ref.~\cite{Buckley:2014fba} based on the simplified model introducing a mediator in the $s$-channel. 
The CMS~\cite{Khachatryan:2016reg} and ATLAS~\cite{Aad:2014vea} collaborations have searched this signature based on ${\cal O}_b$.
The results were recasted to give a tentative bound on the pseudoscalar $A$ portal DM model with the narrow width approximation (NWA) by assuming Br$(A \to\bar \chi \chi)=1$~\cite{Berlin:2015wwa}. 
However, in the region of interest at the future colliders, the mediators may have large couplings to both DM and quarks. 
Therefore, its width is expected to play an important role~\cite{Buckley:2014fba,Abercrombie:2015wmb}. 
Until recently, the CMS collaboration~\cite{CMS-PAS-B2G-15-007} has carried out a search for $b$-jet plus MET in the simplified model framework without adopting the NWA.

The 2$b+$MET channel has not been specifically investigated yet,~\footnote{This statement is only true with respect to the $\Phi_1$-like portal models where $b\bar b$ is associately produced. 
Actually, the 2$b$+MET signature has been investigated in other contexts such as sbottom search or di-Higgs search~\cite{Aaboud:2016nwl,CMS:2016wry,Banerjee:2016nzb}. } despite of a brief mention of sensitivity at the 8 TeV LHC using sbottom search data~\cite{Craig:2015jba}. 
The main argument is that the second $b$-jet is usually soft, thus, it is hard to be detected~\cite{Lin:2013sca}. 
However, an extra $b$-jet, once tagged, will be very helpful to suppress backgrounds. 
With two jets in the final state, the razor variables~\cite{Rogan:2010kb,Chatrchyan:2011ek,CMS:2011eta,Fox:2012ee} will be powerful discriminators at hand. 
This paper is devoted to detailedly analyzing the prospects for the 2$b+$MET channel within the framework of $\Phi_1$-like portal model, with the full consideration of the width of mediator. 
We find that in the region of relatively low mediator mass (below a few hundred GeV), the 2$b$-tagged razor analysis has comparable sensitivity with the mono-$b$ search, if we require the signal to background ratio to be great than a few percent.

This work is organized as follows. 
In Section~\ref{sec:model}, we briefly describe the simplified model for the $\Phi_1$-like portal DM, where the couplings between the heavy mediators and the bottom quarks follow the type-II 2HDM. 
In Section~\ref{sec:collider}, we analyze the LHC searches for the mono-$b+$MET and the $2b+$MET channels.
Particularly, we highlight the possibilities of looking for the $2b+$MET channel by using the shapes of the razor variables as powerful discriminators.
The main results by either using the mono-$b+$MET channel or using the $2b+$MET channel with the aid of the shape analysis are presented in Section~\ref{sec:results}. 
The conclusion is given in Section~\ref{sec:conclusion}.

\section{The simplified model for $\Phi_1$-like portal DM  }
\label{sec:model}

In this section, we construct the simplified model for the $\Phi_1$-like portal DM. 
In the type-II 2HDM, there are two additional neutral Higgs bosons, one $CP$-even $H$ and one $CP$-odd $A$. Their couplings to the down-type quarks and leptons are enhanced by large $\tan\beta$, which is defined as the ratio between vacuum expectation values of $\Phi_2$ and $\Phi_1$. 
The large $\tan\beta$ inputs lead to large production cross sections of $H/A$ associated with $b\bar b$. Moreover, an additional singlet scalar $S$, to which DM couples, may be also presented in the Higgs sector and it has the potential to have a large mixing with $\Phi_1$, thus inheriting features of $\Phi_1$. 
The resulting portal is dubbed $\Phi_1$-like. 
In this paper, we focus on the case that DM is a Majorana fermion $\chi$. 
The corresponding simplified model for the $H$ mediator is
\begin{align}\label{SIM}
-{\cal L}_\chi=&\f{m_\chi}{2}\bar\chi\chi+\f{m_H^2}{2}H^2+ Y_\chi H\bar\chi\chi+Y_bH\bar bb.
\end{align} 
In the class of type-II 2HDM-like models with an additional Higgs singlet of $S$, one has $Y_b=\f{m_b}{v} \tan\beta \times \sin\theta$ with $\sin\theta$ being the mixing factor between $S$ and $\Phi_1$. 
This mixing angle can be sizable, i.e., $\theta\sim \pi/4$. 
Thus, it does not bring a significant suppression.
In other words, $Y_b$ almost follows the doublet-$b$-quark coupling in 2HDM of $\f{m_b}{v} \tan\beta$, with $\tan\beta\gtrsim 10$ of interest in this paper. 
The underlying singlet-DM coupling $Y_\chi$ is not suppressed by any mixing angle and it is assumed to be of order one, which guarantees a substantial branching ratio of $H\ra \chi\chi$. 
If the portal is the $CP$-odd Higgs boson $A$, the simplified model becomes 
\begin{align}
-{\cal L}_\chi=&\f{m_\chi}{2}\bar\chi\chi+\f{m_A^2}{2}A^2+i Y_{\chi}A\bar\chi\gamma_5\chi+iY_bA\bar b\gamma_5b. 
\end{align} 
The Yukawa couplings between $H/A$ and other SM fermions are not explicitly included here, which are assumed to resemble those of the type-II 2HDM. 
In more general sense, $\chi$ may not be a DM particle, while it represents a neutral and stable particle at the collider time scale.

When kinematically allowed, the mediator $H/A$ can decay into bottom quarks and DM particles. Their partial widths are
\begin{align}
\Gamma(H/A \to \chi \chi) &=  \frac{Y_\chi^2 m_{H/A}}{4 \pi} \left (1- \frac{4 m_{\chi}^2}{m^2_{H/A}}\right)^{n/2} ~,~ \label{eq:hxx}\\
\Gamma(H/A \to b \bar b) &=  \frac{3 Y_b^2 m_{H/A}}{8 \pi} \left( 1- \frac{4 m_{b}^2}{m^2_{H/A}}\right)^{n/2}~,~ \label{eq:hbb}
\end{align}
where $n=1$ and 3 for $A$ and $H$, respectively. In principle, the simplified model contains four free parameters, $m_{H/A}$, $m_\chi$ and $Y_b$, $Y_\chi$. For the production mainly through a resonance, $m_\chi$ is almost irrelevant as long as it stays sufficiently small, say $\mathcal{O}(1)$ GeV. We will further take a few samples of $Y_\chi$ to reduce the number of free parameters.


\section{Collider searches}
\label{sec:collider}

\subsection{Preliminary for the signals}

In the simplified model, the DM can be pair produced via the $s$-channel $A$ mediation.~\footnote{The production cross sections via the $H$ and $A$ mediation only differ in percentage level. 
We always focus on the $A$ mediation case throughout our discussions.}
The production cross section for the process can be calculated either in the four-flavor-number (4F) scheme~\cite{Campbell:2004pu,Maltoni:2012pa,Gang:2016vfn} where the leading order (LO) process is $gg \to b \bar{b} A (\to \chi \chi)$, or in the five-flavor-number (5F) scheme where the LO process is $b \bar{b} \to A(\to \chi \chi)$. The calculation in the 5F scheme is highly simplified because of the reduced number of final state particles. In addition, the potentially large logarithms arising from collinear splitting of the initial quarks and gluons have already been resumed in the 5F parton distribution function. 
However, the 4F scheme that takes into account the full kinematics of the final states at the LO, is easier to simulate and will be adopted in this work. 
In the following discussion, the signal events and production cross section are generated by MadGraph5\_aMC@NLO~\cite{Alwall:2014hca}.

To get some idea on the relative production cross section between the signals with single $b$-tagged jet and two $b$-tagged jets, we first generate an inclusive event sample of $gg \to b \bar{b} A$ without applying any cuts to the $b$-jets.~\footnote{For simplicity, we have assumed the signal is on-shell A production with subsequent decay to DM particles and the coupling $Y_b$=1.} 
The inclusive production cross sections for different $A$ masses are given in the second column of Tab.~\ref{xsec}. 
Next, we require at least one or two $b$-jets in the final state that have $p_T > 20$ GeV and $|\eta|<2.5$. 
The fraction of events that pass these conditions are recorded in the third and fourth column of the same table. 
Depending on the Higgs mediator mass, the rates of signal with two $b$-jets are around 4-8 times smaller than the rates of signal with one $b$-jet. 
On the other hand, a large MET, i.e. large transverse momentum of Higgs is required in the mono-$b$ search. 
The efficiency for a tentative cut on $p_T(A)$ is provided in the last column of the table. 
This cut will lead to lower signal rate than two $b$-jets cut at low $m_A$ and higher signal rate at high $m_A$ region. We find our results in Tab.~\ref{xsec} well match those given in Ref.~\cite{Wiesemann:2014ioa}.

 \begin{table}[htb]
\begin{center}
  \begin{tabular}{|c|c|c|c|c|c|c|c|c|}\hline
 $m_A$ (GeV) & $\sigma^{\text{incl}}$ (pb) & $\epsilon(\geq 1 j_b)$ & $\epsilon(\geq 2 j_b)$ & $\epsilon(p_T(A)>100~\text{GeV})$ \\ \hline \hline
 125 &  1562  & 0.374 & 0.0472 & 0.0257 \\
 500 &  6.83  & 0.602 & 0.131 & 0.177 \\
 1000 &  0.2115  & 0.662 & 0.170 & 0.298\\
 2000  & 0.002748 & 0.696 & 0.199 & 0.409 \\   \hline
  \end{tabular}
  \caption{\label{xsec} Cross sections and cut efficiencies of mono-$b$ and 2$b$-jets signatures at 14 TeV LHC, where $Y_b=1$. The $b$-jets satisfy $p_T > 20$ GeV and $|\eta|<2.5$}
\end{center}
 \end{table}

\subsection{ Mediator: Narrow Width (NW) versus Wide Width (WW) }

In the previous studies, such as in Ref.~\cite{Berlin:2015wwa}, the mediator is usually assumed to resume the NWA. 
However, in a large parameter space that can be explored at the LHC, the NWA tends to be invalid to certain degree, which causes significant errors. 
The reason is presented as follows. 
To guarantee sufficient signal production rate at the LHC via the process $gg\rightarrow bbA(\rightarrow \chi\bar\chi)$, both $Y_b$ and $Y_\chi$ should be of order one.
As will be shown later, even the high-luminosity LHC (HL-LHC) can only probe the parameter regions of $Y_b \gtrsim 0.2$ for $m_{A} \gtrsim 100$ GeV, and an even much larger $Y_b$ is required for the heavier ${A}$. 
Consequently, the typical decay width is such that $\Gamma (A\to \chi\bar\chi+b\bar b) \gtrsim 0.1 m_{A}$, which apparently violates the NWA. 
In the present work, we will still incorporate the NW scenario for the purposes of validation as well as comparison. 
Note that in the WW scenario when $Y_b$ is sufficiently large, the event rate of $\sigma (gg \to b \bar{b} A(\to \chi \chi))$ will become insensitive to $Y_b$, because the $Y_b^2$ factor in the production is cancelled by the one from $\Gamma(A \to b \bar{b})$, which dominates the propagator factor when one considers the regions close to the resonance. 
Such a behavior will be explicit in exploring the LHC sensitivity to the simplified model on the $m_{A}-Y_b$ plane.

In the LHC studies, the Monte Carlo (MC) events for the signals and backgrounds are generated at the LO by MadGraph5\_aMC@NLO~\cite{Alwall:2014hca}, where the signal cross section is also calculated. 
The Pythia6~\cite{Sjostrand:2006za} is used for decaying the SM particles, parton showering and hadronization. The hadron-level events are passed through Delphes3~\cite{deFavereau:2013fsa} with the default ATLAS setup to simulate the detector effects. 
As in the ATLAS analysis~\cite{Aad:2014vea}, we set the $b$-tagging efficiency to be 60\,\%, and the corresponding mis-tagging rates for the charm- and light-flavor jets are 0.15 and 0.008, respectively.~\footnote{
According to Ref.~\cite{ATL-PHYS-PUB-2015-022}, the $b$-tagging rate can reach $\sim 75\,\%$ while keeping the rejections rates intact at the LHC run-2.
This is due to the additional insertable $b$-layer.
In this work, the LHC $b$-tagging efficiency for LHC Run-1 is used for both extrapolated mono-$b$ analysis and razor analysis in order to have a more conservative comparison.} 
Throughout the simulation, we find that the difference between the scalar and the pseudoscalar is small. 
Therefore, we will focus on the pseudoscalar in our following study, which is also motivated by DM phenomenology~\cite{Guo:2014gra}.

\subsection{Recast ATLAS mono-$b$ jet analysis}
\label{sec:monob}

Even though our signal contains two $b$-quarks at the parton level, the second $b$-jet is usually too soft to be tagged, as shown in Tab.~\ref{xsec}. 
In this case, the final state is featured by a single $b$-jet and large MET, i.e., the mono-$b$ signature. 
A dedicated search for the signal of heavy quark associated DM pair production has been carried out by ATLAS collaboration~\cite{Aad:2014vea} at 8 TeV with the integrated luminosity of 20.3 fb$^{-1}$. 
In their analysis, the signal region SR1 especially focus on the mono-$b$ signature, which requires lepton veto ($N_\ell=0$), large MET ($\slashed{E}_T> 300$ GeV), energetic $b$-tagged jet ($p_T(b_1)>100$ GeV), low jet multiplicity ($N_j= $1-2) and large azimuthal angle separation between jets and MET ($\Delta \phi_{\min} (j_i,\slashed{E}_T) >1.0$). 
However, their results are only presented in terms of effective operators. 
In this subsection, we will recast their analysis in our simplified models.

We choose eight benchmark points, as shown in Tab.~\ref{tab:eff8b}. 
For each benchmark point, 10$^5$ signal events are generated at the parton level, with a cut $p_T(b_1)>50$ GeV on the leading $b$-jet. 
Then, we apply the cuts of the signal region SR1 and record the number of the remnant events in Tab.~\ref{tab:eff8b}. 
It turns out that the cut efficiencies increase with the increasing mediator masses in the NW scenario. 
This is well expected, since the final states become more energetic as the mediator becomes heavier. 
By contrast, the efficiency increases more slowly or even decreases in the high mass region in the WW scenario, where the off-shell contribution is dominant.

\begin{table}[htb]
\center
  \begin{tabular}{|c|c|c|c|c|c|c|c|c|}
  \hline
  $m_A$(GeV) & 100 & 200 & 300 & 500 & 700 & 1000 & 1500 & 2000 \\ \hline\hline
$ \epsilon^{\text{NW}}_{\text{SR1}}/10^{5}$ & 21 & 72 & 164 & 395 & 642 & 944 & 1331 & 1510  \\
$ \epsilon^{\text{WW}}_{\text{SR1}}/10^{5}$ & 19 & 76 & 152 & 336 & 459 & 587 & 599 & 441 \\ \hline
  \end{tabular}
\caption{\label{tab:eff8b} 
Cut efficiencies of the signal region SR1 in the ATLAS search in the NW scenario (upper row) and WW scenario (lower row). }
 \end{table}

\begin{figure}[htb]
\includegraphics[width=0.45\textwidth]{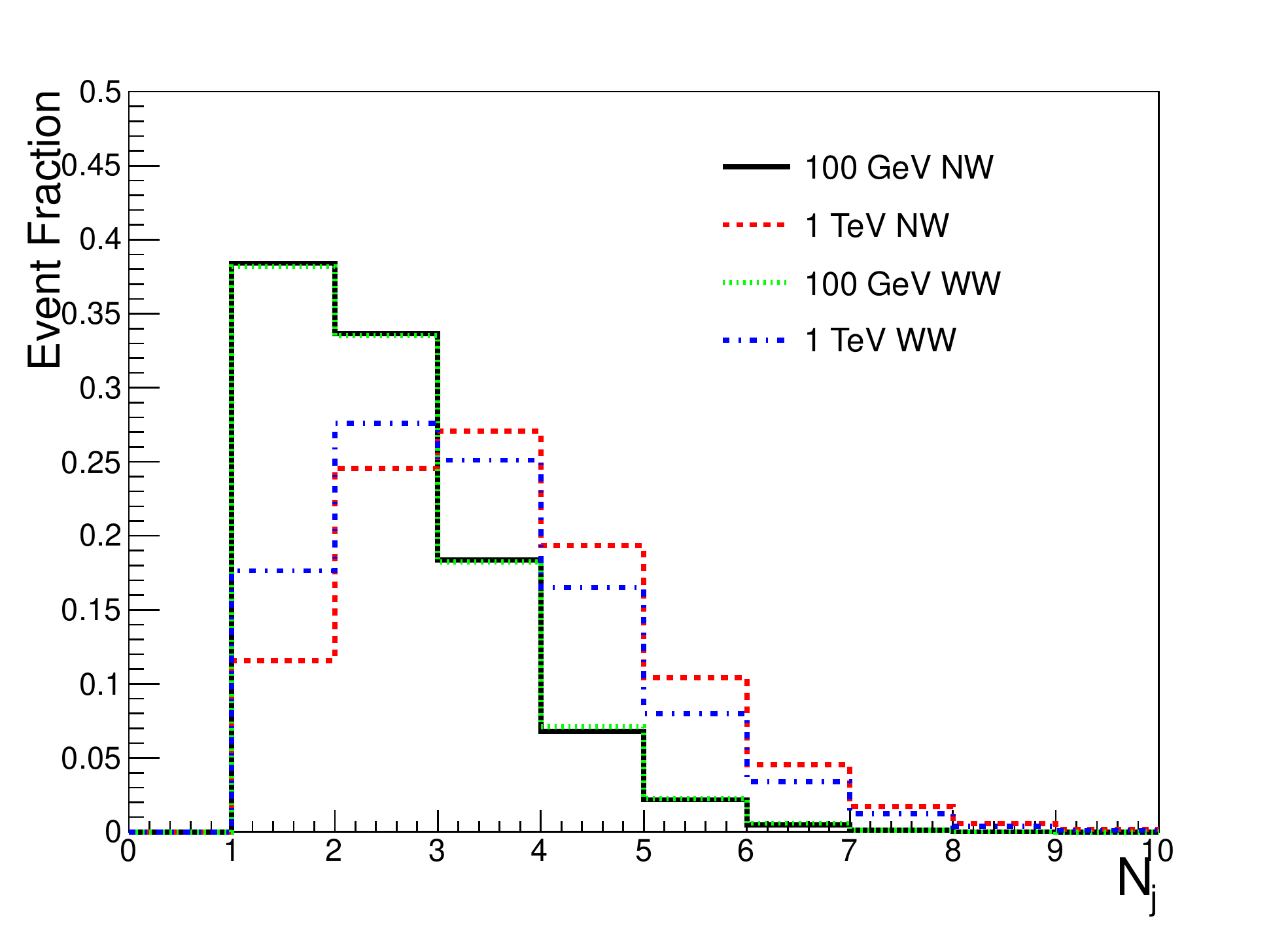}
\includegraphics[width=0.45\textwidth]{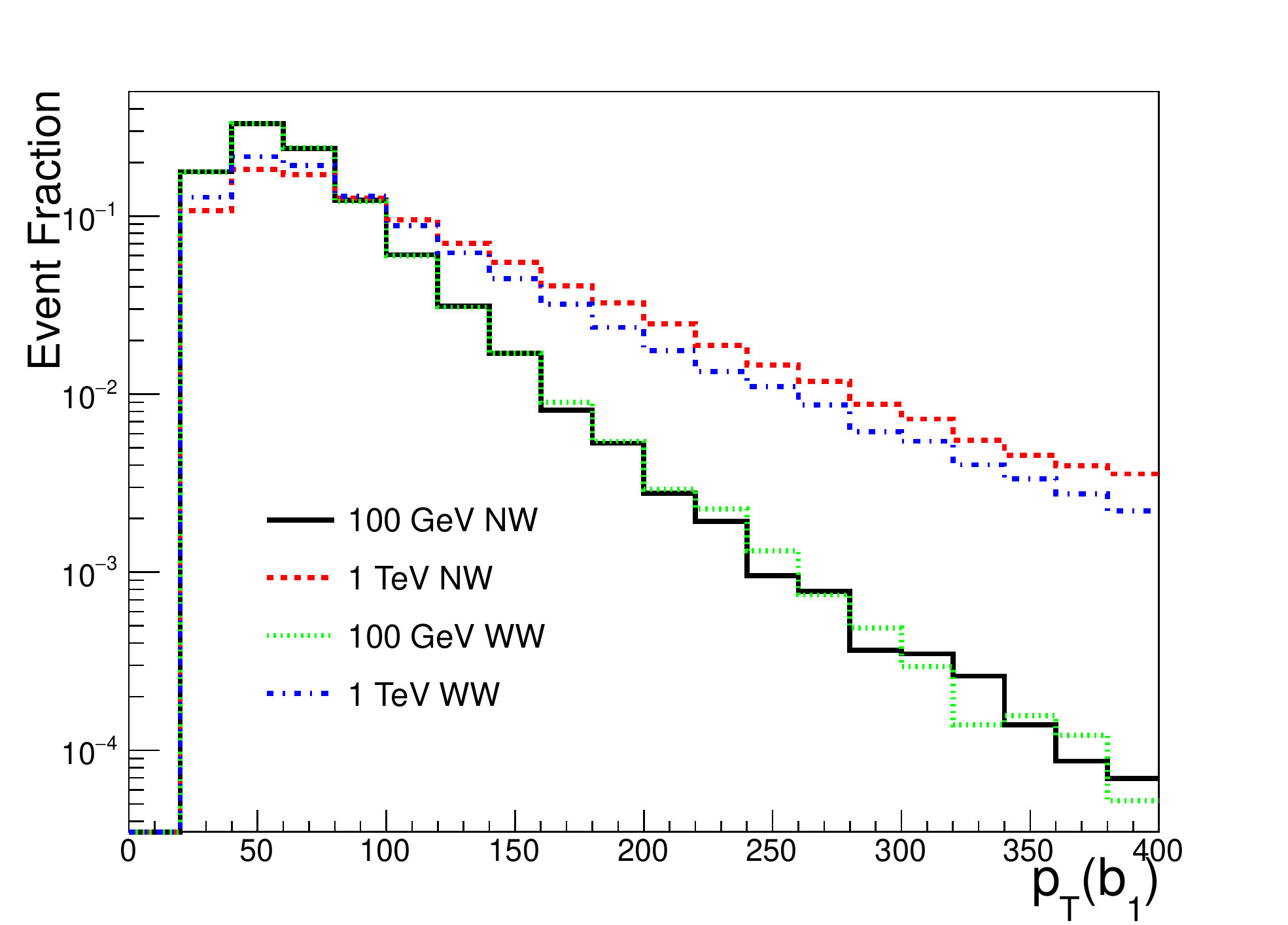} \\
\includegraphics[width=0.45\textwidth]{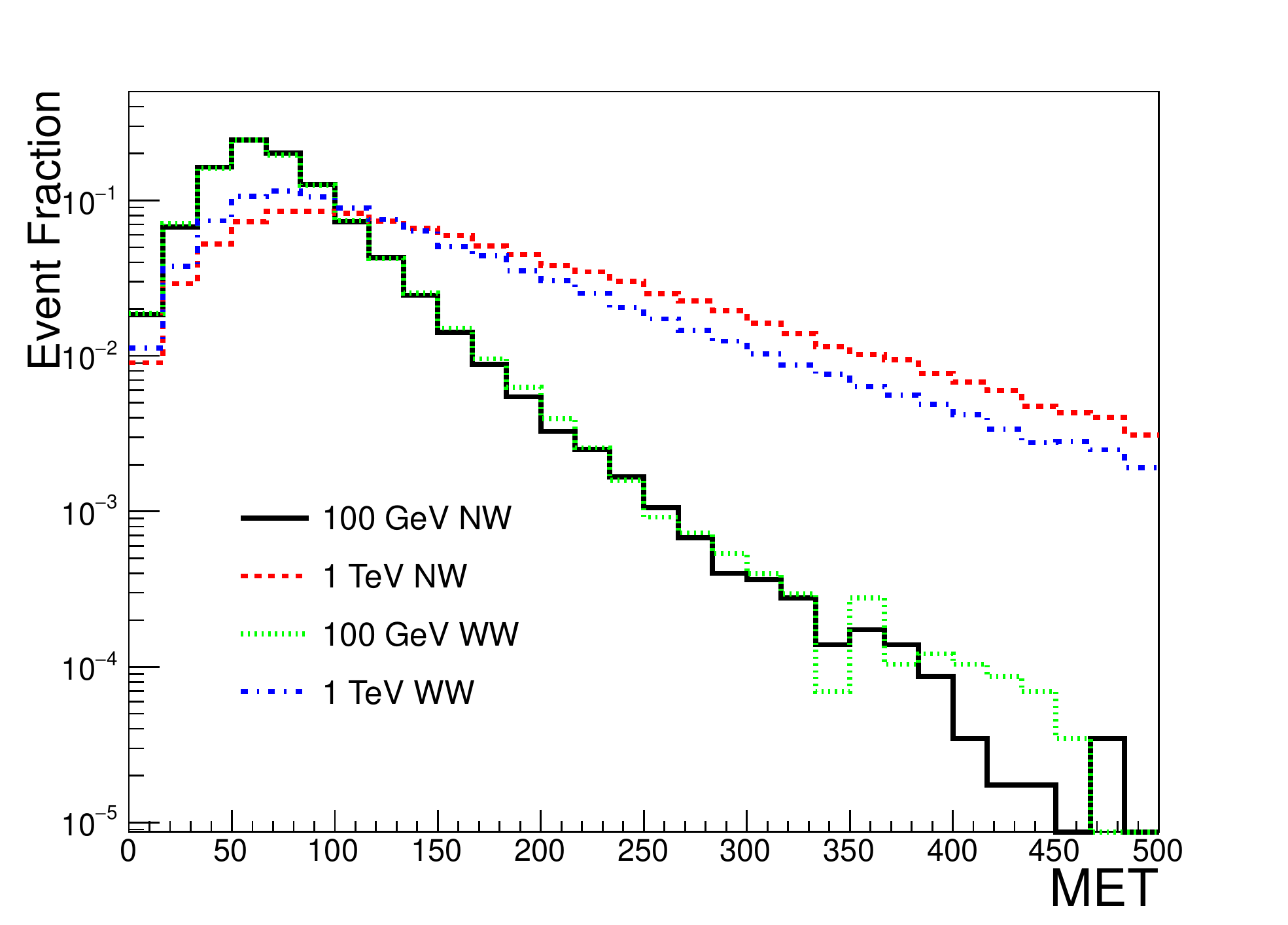}
\includegraphics[width=0.45\textwidth]{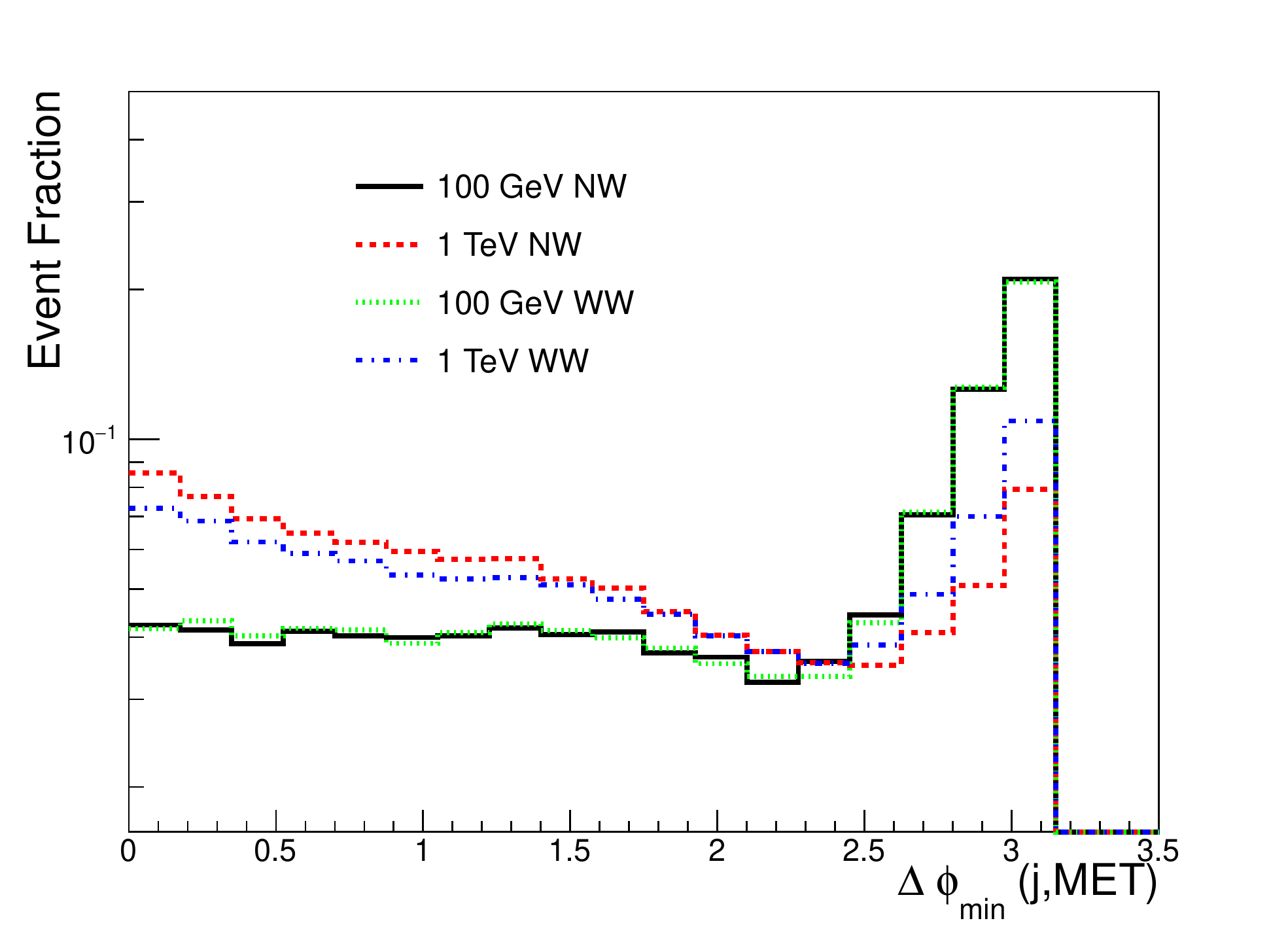}
\caption{\label{kins} The distributions of $N_j$, $p_T(b_1)$, MET and $\Delta \phi_{\min} (j_i,\slashed{E}_T)$ for the NW and WW scenarios with mediator masses of 100 GeV and 1 TeV at 8 TeV LHC.  }
\end{figure}

To have a closer look at the above features, we plot the distributions of the discriminant variables in Fig.~\ref{kins} for both scenarios prior to any selection cuts. 
Two benchmark points have been chosen to represent the low-mass region and the high-mass region, respectively. 
When the mediator is light, the differences between the NW and WW scenarios are negligible for all variables. While in the heavy mediator region, the overall energy scale of the final states becomes higher and even the second $b$-jet is likely to be tagged.
Nevertheless, the significant off-shell mediator contribution in the WW scenario softens the final state, thus rendering softer $p_T(b_1)$ and MET and less $N_j$. 
In addition, the distribution of $\Delta \phi_{\min} (j_i,\slashed{E}_T)$ is also affected by the energy scale of the final states, namely, a higher energy scale leads to more QCD radiation thus a smaller $\Delta \phi_{\min} (j_i,\slashed{E}_T)$. 

There are 440 observed events in the signal region SR1 with expected $385\pm 35$ SM background events, which can yield an upper limit on the new physics cross section of $\sim6.1$ fb at the 95\% confidence level (CL). 
The corresponding limit on the production cross sections for the benchmark points can be directly calculated via $\sigma_S^8 = 6.1~\text{fb}/\epsilon_{\text{SR1}}$, where the values of $\epsilon_{\text{SR1}}$ can be simply derived from Tab.~\ref{tab:eff8b}. 

In order to make a tentative estimation on the future discovery prospects of this channel, we extrapolate the 95\% exclusion bound obtained at 8 TeV with two following assumptions as adopted in Refs.~\cite{CMS:2013xfa,Buchmueller:2015uqa}: (1) the signal and background efficiencies at different collider energies remain unchanged; (2) the uncertainty of background is scaled by $\sqrt{B}$ where $B$ is the total number of background events after the selection. 
The corresponding 95\% exclusion limit at the 14 TeV LHC, $\sigma^{14}_S$, can be extrapolated as
\begin{align}
\sigma^{14}_S=\sqrt{\frac{\sigma^{14}_B}{\sigma_B^8}} \sqrt{\frac{\mathcal{L}^8}{\mathcal{L}^{14}}} \sigma_S^8\,,
\end{align}
where $\sigma_B$ is the background production cross section and $\mathcal{L}$ is the integrated luminosity, with the superscripts indicating the LHC center-of-mass energy.

In the simplified model, the mono-$b$ signal production cross sections in the NW and WW scenarios can be calculated via the following relations
\begin{align}
\sigma^{\text{NW}}(Y_b) = &Y^2_b \times \sigma^{\text{NW}}(Y_b=1),\\
\sigma^{\text{WW}}(Y_b) =& \frac{Y^2_b}{\frac{2}{5} + \frac{3}{5}Y^2_b} \times \sigma^{\text{WW}}(Y_b=1),
\end{align}
where we have chosen $Y_\chi=1$ and the cross sections (for a given benchmark point) with the fixed input of $Y_b=1$ are calculated by MadGraph5\_aMC@NLO and tabulated in Tab.~\ref{tab:xsecb}. 
With the signal cross section, a benchmark point will be excluded by the mono-$b$ search at 95\% CL if its production cross section given in Tab.~\ref{tab:xsecb} is larger than the corresponding $\sigma^{8/14}_S$ as calculated above. 
For comparison (with the $2b$-channel), we postpone the presentations of results (8 TeV bound and 14 TeV prospect) to Section~\ref{sec:results}.

\begin{table}[htb]
\center
  \begin{tabular}{|c|c|c|c|c|c|c|c|c|c|}
  \hline
  \multicolumn{2}{|c|}{$m_A$(GeV)} & 100 & 200 & 300 & 500 & 700 & 1000 & 1500 & 2000 \\ \hline \hline
 \multirow{2}{*}{8 TeV}& $ \sigma^{\text{NW}}_{p_T(b_1)>50~\text{GeV}}$ (pb) & 89.95 & 15.01 & 3.75 & 0.45 & 0.082 & 0.010 &  $5.6 \times 10^{-4}$ & 4.26 $\times 10^{-5}$  \\
& $ \sigma^{\text{WW}}_{p_T(b_1)>50~\text{GeV}}$ (pb) & 32.46 & 5.47 & 1.43 & 0.19 & 0.04 & $7.04\times 10^{-3}$ & $8.86\times 10 ^{-4}$ & $2.2 \times 10^{-4}$ \\ \hline
 \multirow{2}{*}{14 TeV} &$ \sigma^{\text{NW}}_{p_T(b_1)>50~\text{GeV}}$ (pb) & 310.4 & 61.1 & 17.6 & 2.68 & 0.62 & 0.11 & 0.011 & 0.0017  \\
& $ \sigma^{\text{WW}}_{p_T(b_1)>50~\text{GeV}}$ (pb) & 112.8 & 22.3 & 6.55 & 1.06 & 0.27 & 0.055 & 0.0078 & 0.0018 \\ \hline
  \end{tabular}
\caption{\label{tab:xsecb} Production cross section for NW scenario and WW scenario at the LHC.  We have set $Y_b$=1 and the leading $b$-jet in the final state is required to have $p_T(b_1)>50$ GeV. }
 \end{table}

\subsection{Shape analysis with two $b$-jets}
\label{sec:razor}

As the central topic of this paper, we investigate the abandoned channel of $2b$ plus MET in this subsection. 
Comparing to the single $b$-jet plus MET channel, it has an additional suppression factor of 4-8 on its production cross section, as shown in Tab.~\ref{xsec}. 
Nevertheless, the second $b$-jet in the final state may serve as another handle to  reduce the SM background, which improves the signal significance in turn. 
In particular, the shapes of razor variables are found to be powerful discriminators.

While generating the signal events, both $b$-jets are required to have $p_T(b_i)>20$ GeV, $|\eta(b_i)|<2.5$ and $\Delta(b_1,b_2) >0.4$ at the parton level. 
At the LHC detectors, the charged leptons (in particular from cascade decays) may be missed and the light flavor jets may be mis-tagged as $b$ jets. 
Therefore, they give rise to the main SM backgrounds~\footnote{We require at least two jets with $p_T(j_i)>20$ GeV, $|\eta(j_i)|<2.5$ and $\Delta(j_1,j_2) >0.4$ at parton level for all backgrounds except for $t\bar{t}$.} for the signal, $t\bar{t}$, QCD multijets, $W(\to \ell \nu)+$jets and $Z(\to \nu \nu)+$jets. 
The NNLO $t\bar{t}$ production cross section of $\sigma(t\bar{t}) = 920$ pb~\cite{Kidonakis:2011zn} is used in our analysis. 
Since the higher order correction tends to reduce the production cross sections of QCD multi-jet~\cite{Badger:2012pf}, $W(\to \ell \nu)+$jet and $Z(\to \nu \nu)+$jets~\cite{Campbell:2003hd} in the region with high jet multiplicity, the LO cross sections calculated by MadGraph5\_aMC@NLO are chosen to make a conservative evaluation: $\sigma(\text{QCD}) = 3.4 \times10^{7}$ pb, $\sigma(W(\to \ell \nu)jj) = 3360$ pb and  $\sigma(Z(\to \nu \nu)jj) = 714$ pb.

At the detector level, the jet candidates are reconstructed by the anti-$k_t$ jet algorithm with the radius parameter of $R=0.4$ in the FastJet~\cite{Cacciari:2011ma}. 
Only events with at least two central jets are selected for the later analysis. 
The central jets should have $p_{T}>40$ GeV, $|\eta|<2.5$ and $\Delta R >0.4$ away from other jets. 
In the case of more than two central jets in the final state, all $n$ jets are partitioned into two group (dubbed megajet) with ($2^{n-1}-1$) possible ways. 
The megajet momentum is defined by the vector sum of all jets momenta in each group. 
The partition that minimizes the sum of two megajets invariant mass square is chosen, and the corresponding two megajets are denoted by $J_{1,2}$. The razor variables~\cite{Rogan:2010kb,Chatrchyan:2011ek,Fox:2012ee} are defined as 
\begin{align}
M_R &\equiv \sqrt{(E(J_1) +E(J_2))^2 - (p_z(J_1)+p_z(J_2))^2}~, \quad 
R\equiv \frac{M^T_R}{M_R}~,
\end{align}
with
\begin{align}
M_R^T\equiv \sqrt{\frac{\slashed{E}_T (p_T(J_1)+ p_T(J_2)) - \vec{\slashed{E}}_T \cdot (\vec{p}_T(J_1) + \vec{p}_T(J_2) ) }{2}}~.
\end{align}
The variable $M_R$ provides an estimation on the energy scale of a certain process.
Thus, the signal process involving heavy particles typically has larger $M_R$ than the background processes. 
The variable $R^2$ is correlated with the angular separation between the $J_1$ and $J_2$. 
In the background processes where two megajets are nearly back-to-back, $R^2$ is close to 0; whereas in the signal process especially when the DM particles carry away large energy, the $R^2$ variable tends to be fairly sizable.

\begin{figure}[htb]
\includegraphics[width=0.45\textwidth]{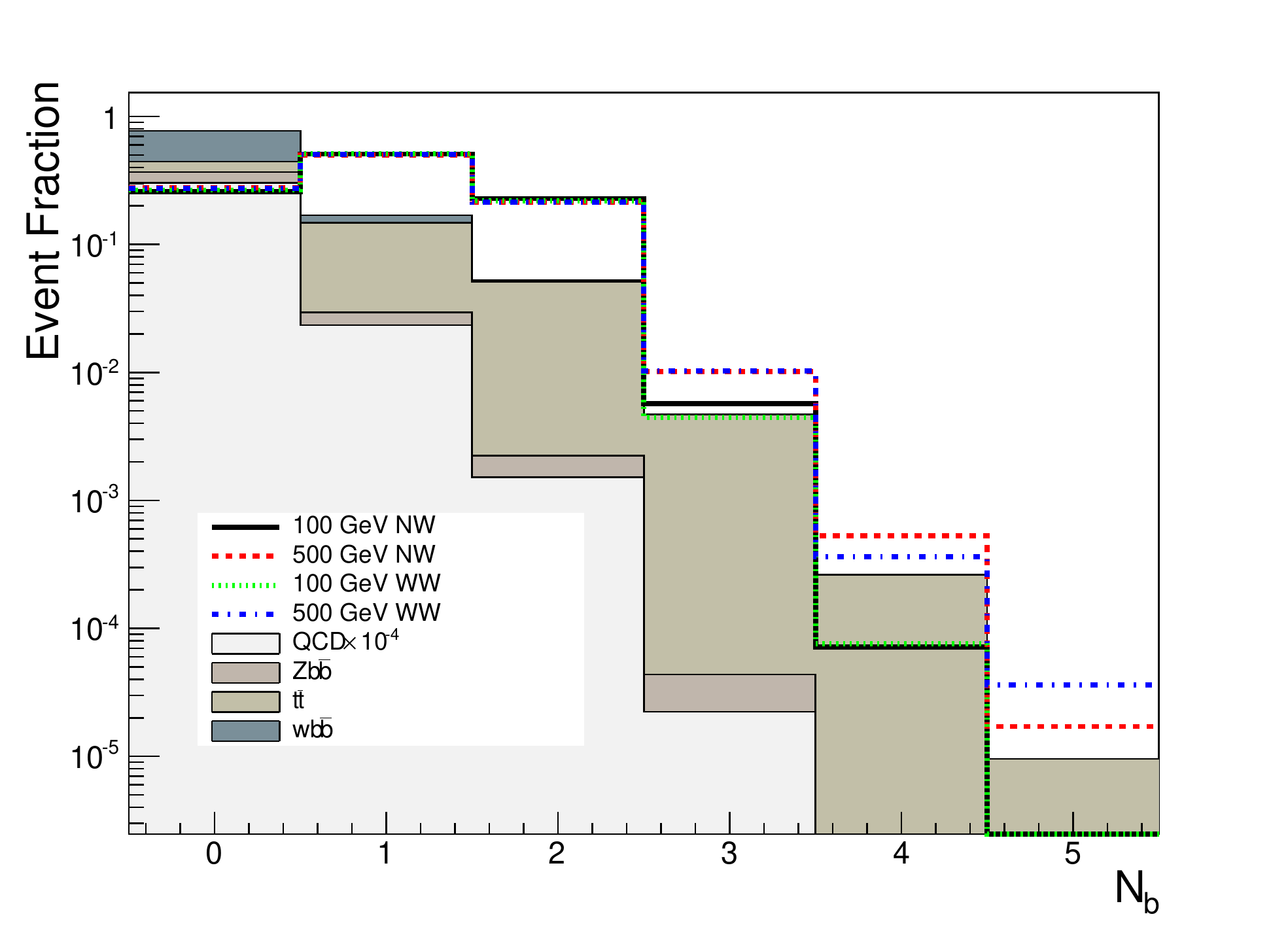}
\includegraphics[width=0.45\textwidth]{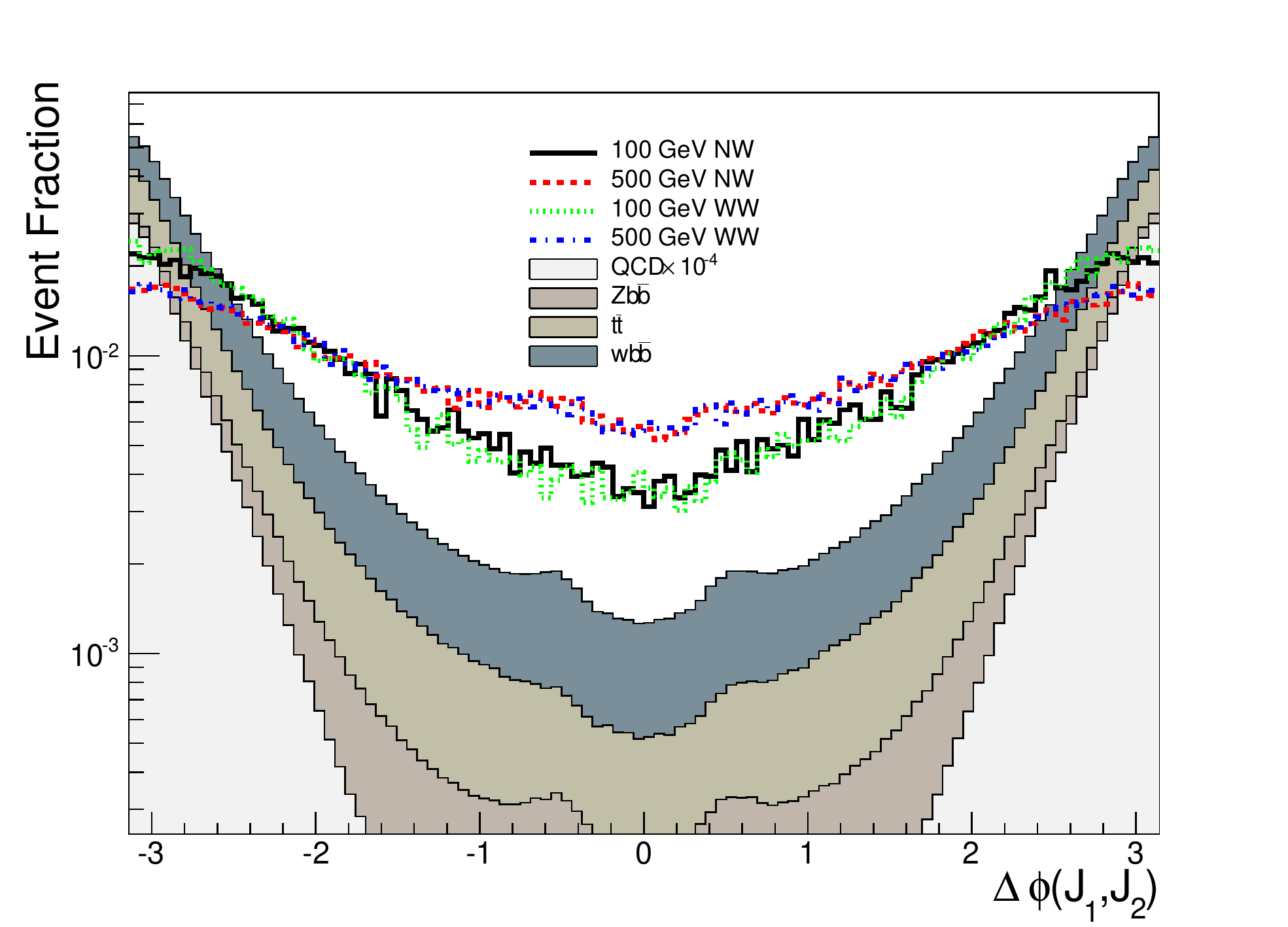} \\
\includegraphics[width=0.45\textwidth]{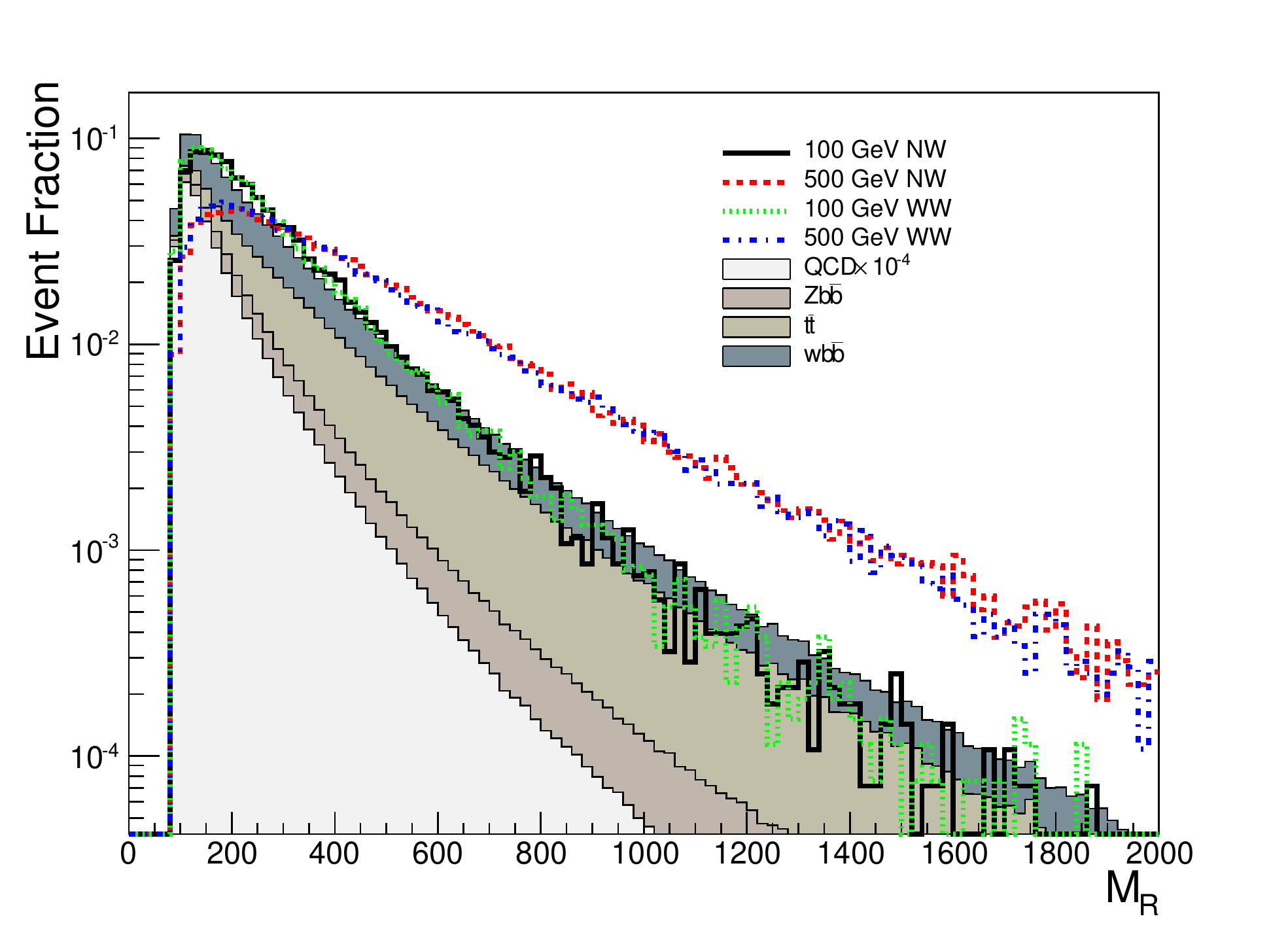} 
\includegraphics[width=0.45\textwidth]{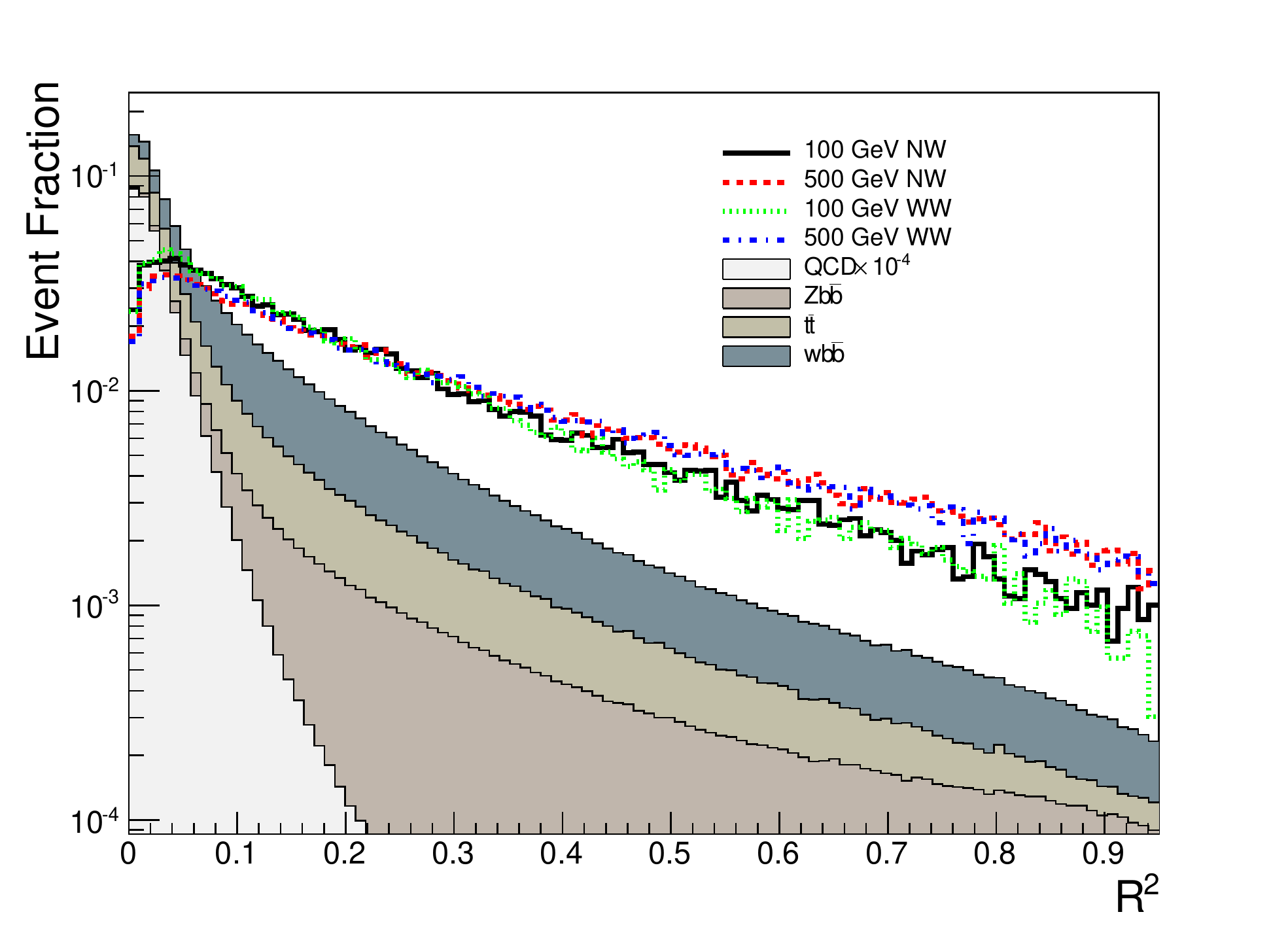}
\caption{\label{kinsbb} The distributions of $N_b$, $M_R$, $R^2$ and $\Delta \phi (J_1,J_2)$ for NW and WW scenarios with mediator mass 100 GeV and 500 GeV at 14 TeV LHC. More explanations for the relative size of distributions in backgrounds can be found in text. }
\end{figure}

Besides of the razor variables, we find that the number of $b$-tagged jets $N_b$ and the azimuthal angle separation between two megajets $\Delta \phi(J_1,J_2)$ also play important roles in separating the signal and background events. 
We demonstrate their distributions in Fig~\ref{kinsbb}, using the events with at least two central jets. 
In the figures, all the background distributions have been added up with weights proportional to their production cross sections except for the QCD background, whose weight is reduced by a factor of $10^4$ to maintain the features of other backgrounds. 
The distributions for the summed background and each signal have been normalized to one.
Several observations are available:
\begin{enumerate}

\item The signals have higher $b$-jet multiplicity than the backgrounds, with peak at $N_b =1\sim 2$. In the high $b$-jet multiplicity region, the backgrounds are dominated by $t\bar{t}$. 
Additionally, the $N_b$ distributions of signal processes only have a weak dependence on the mediator mass.

\item The azimuthal angles between two megajets of signals are much smaller than those of the backgrounds. 
The signals involving a heavier mediator have smaller azimuthal angle separations, because two megajets are recoiled with higher energy for a heavier mediator, thus closer in azimuthal angle. There is a singular behavior at around $\Delta \phi \sim 0.4$, due to the requirement of angular separation $\Delta R>0.4$ between jets.

\item For mediator mass of $m_A= 100$ GeV, the signal has similar $M_R$ distribution with the backgrounds, since the typical mass scale is close to the electroweak scale. Nevertheless, $M_R$ increases substantially with the larger mediator mass of $m_A= 500$ GeV.

\item On the other hand, the $R^2$ distribution for the signal with $m_A= 100$ GeV, due to its much larger MET, is already much harder than that for the backgrounds. Meanwhile, the increase of $R^2$ with the mediator mass increasing from 100 GeV to 500 GeV is mild.

\end{enumerate}
We can also see from the plots that in the mass region of interest ($m_{A} \lesssim 500$ GeV), the kinematic difference between the NW scenario and WW scenario is small.

As in the CMS experiment, events used for shape analysis could be collected with a trigger based on a loose selection cuts on $M_R$ and $R^2$. Since strong cuts on $M_R$ and $R^2$ will be applied in our following analysis, we do not consider the trigger efficiency at this stage. 
The following preselection cuts are applied to the signal events before carrying out a dedicated shape analysis: (i) no isolated electron or muon; (ii) $|\Delta \phi(J_1,J_2)|<2.5$; (iii) exactly one $b$-tagged anti-$k_t$ jet in each megajet; (iv) $M_R >300$ GeV and $R^2 > 0.1$. 
The signal selection efficiencies with and without the $b$-tagging cut are shown in Tab.~\ref{tab:selec14bb}. 
It is seen that, the selection efficiency increases substantially with the increasing mediator mass; the WW scenario has lower efficiency than the NW scenario with same mediator mass because of the off-shell contribution discussed before, and their difference becomes more significant in the higher mediator mass region. 
The $b$-tagging efficiency is $\sim 20\%$ in the full mediator mass region. 

\begin{table}[htb]
\center
  \begin{tabular}{|c|c|c|c|c|c|c|c|c|} \hline
  $m_A$(GeV) & 100 & 200 & 300 & 500 & 700 & 1000 & 1500 & 2000 \\ \hline \hline
$ \epsilon^{\text{NW}}_{\text{Pre}}/10^{6}$ & 2154 & 4535 & 7768 & 14360 & 19346 & 25610 & 31818 & 35552 \\
$ \epsilon^{\text{NW}}_{\text{$b$-tag}}/10^{6}$ & 502 & 966 & 1464 & 2656 & 3573 & 4813 & 5941 & 6739 \\
$ \epsilon^{\text{WW}}_{\text{Pre}}/10^{6}$ & 1834 & 4509 & 7368 & 12751 & 16667 & 20646 & 22091 & 21586  \\
$ \epsilon^{\text{WW}}_{\text{$b$-tag}}/10^{6}$ & 428 & 933 & 1418 & 2283 & 3020 & 3881 & 4104 & 3933 \\ \hline
  \end{tabular}
\caption{\label{tab:selec14bb} The preselection efficiencies with and without $b$-tagging cut for NW scenario and WW scenario. }
 \end{table}

More refined cuts on the razor variables depend on the shape of background events. 
As we can observe in the lower panels of Fig.~\ref{kinsbb}, the distribution of each background has a simple exponential dependence on $M_R$ and $R^2$ in the region $M_R \gtrsim 200$ GeV and $R^2 \gtrsim 0.1$. 
In contrast to the tails of the $\slashed{E}_T$ distribution which is difficult to model, the distributions of the razor variables over a wide range can be well described by a probability function with  two exponential components~\cite{CMS:2011eta}
\begin{align}
P(R^2, M_R) = f \times e^{-k(M_R-x_0)(R^2-y_0)} + (1-f) e^{-k' (M_R-x'_0) (R^2-y'_0)}~. \label{pdf}
\end{align}
This is especially helpful for the QCD background, because it is important but has too large production cross section to simulate sufficiently. 
The clean shape of razor variables can be used to predict the number of background events at the tail without heavy use of MC simulations. 
Furthermore, it can be found that the shape of the two dimensional $M_R$-$R^2$ distribution are not biased by the $b$-tagging requirement.

\begin{figure}[htb]
\includegraphics[width=0.24\textwidth]{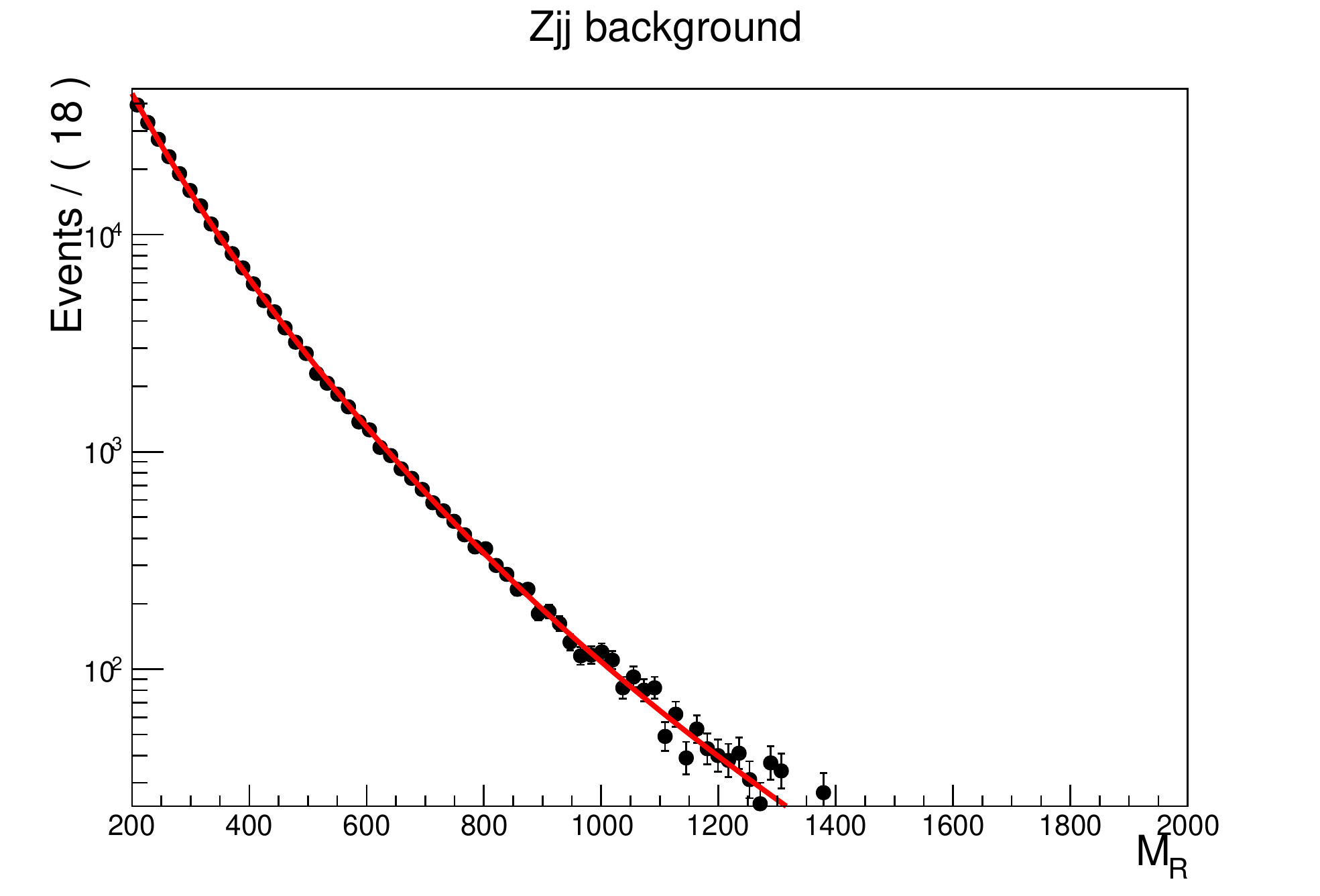}
\includegraphics[width=0.24\textwidth]{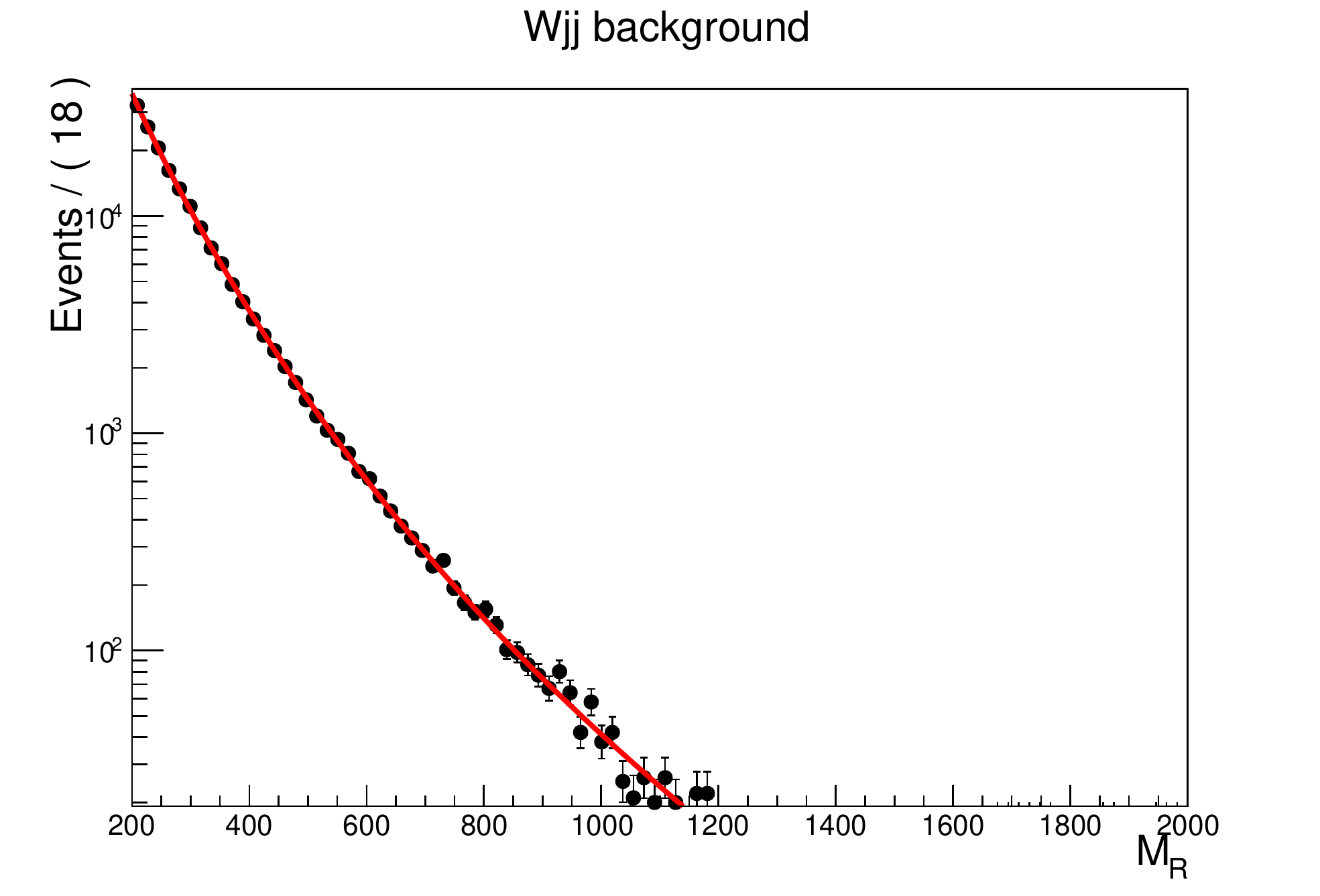} 
\includegraphics[width=0.24\textwidth]{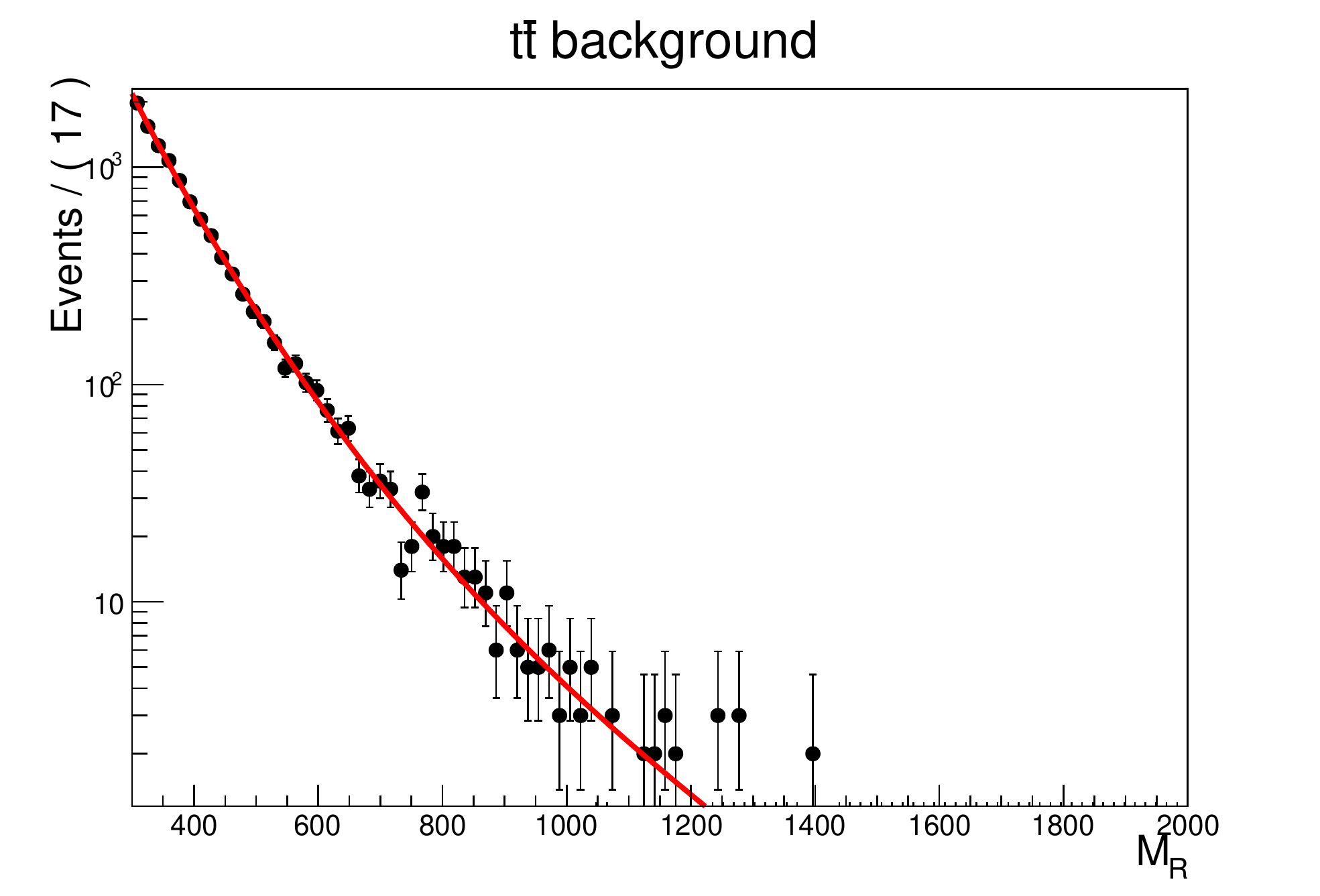}
\includegraphics[width=0.24\textwidth]{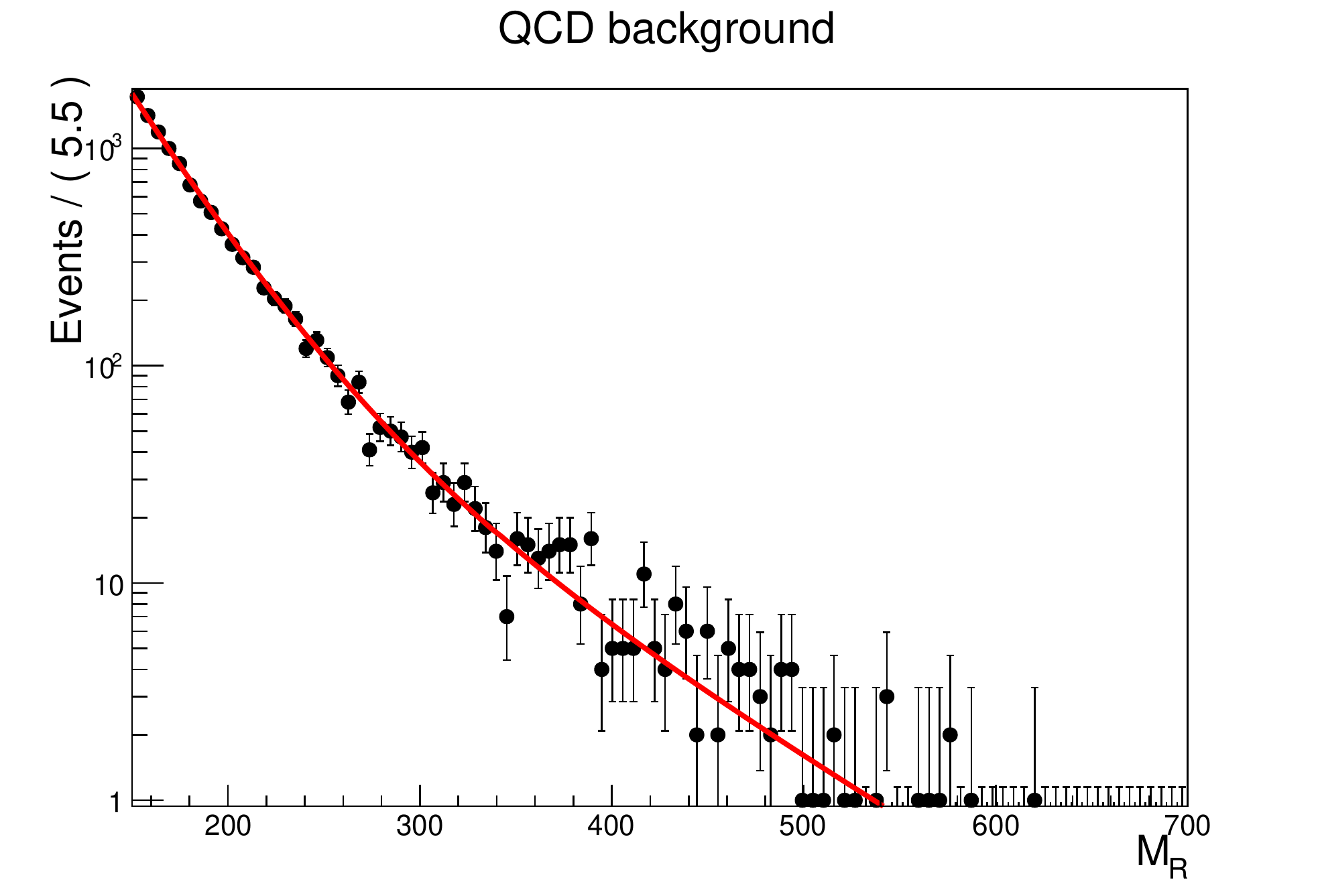} \\
\includegraphics[width=0.24\textwidth]{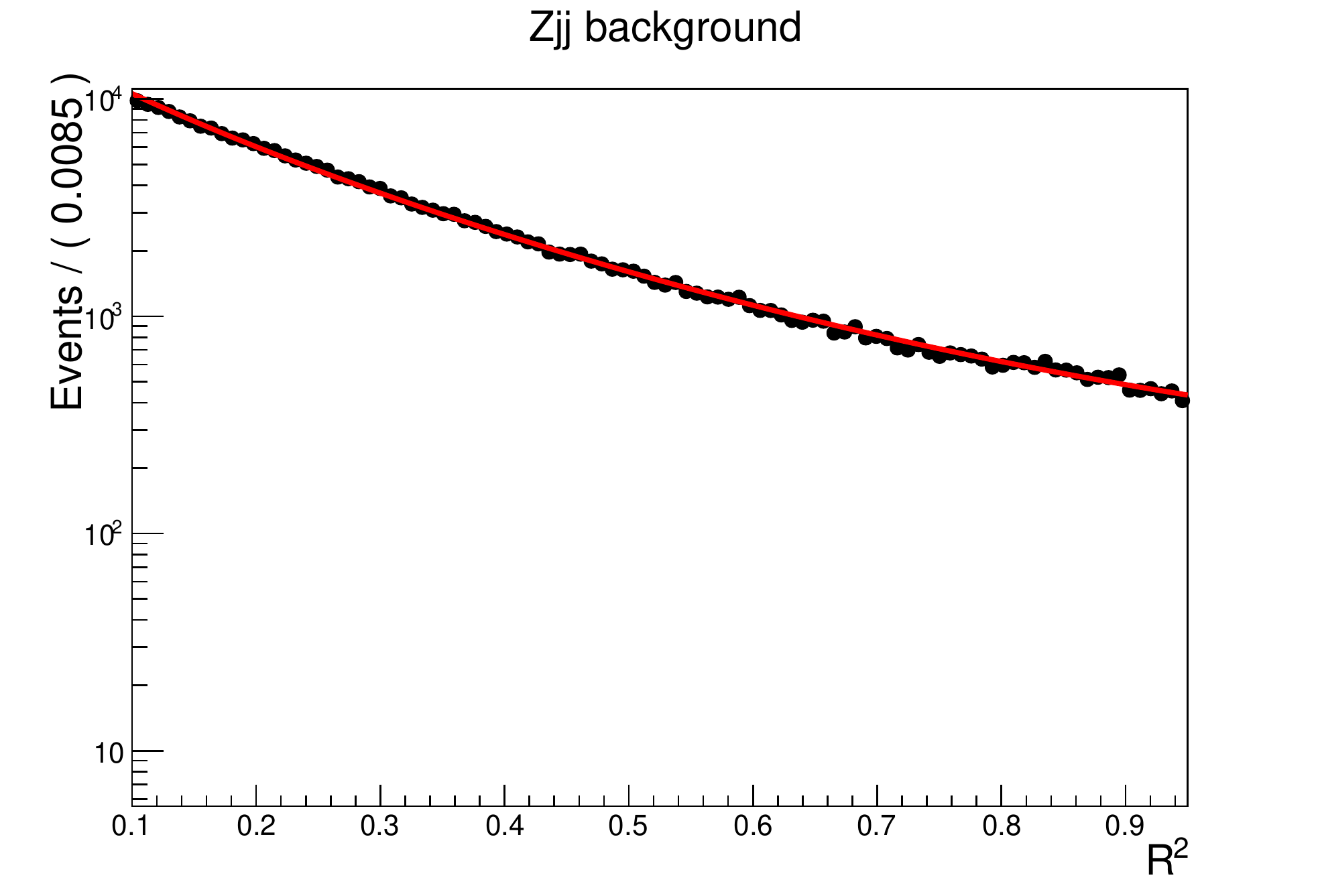}
\includegraphics[width=0.24\textwidth]{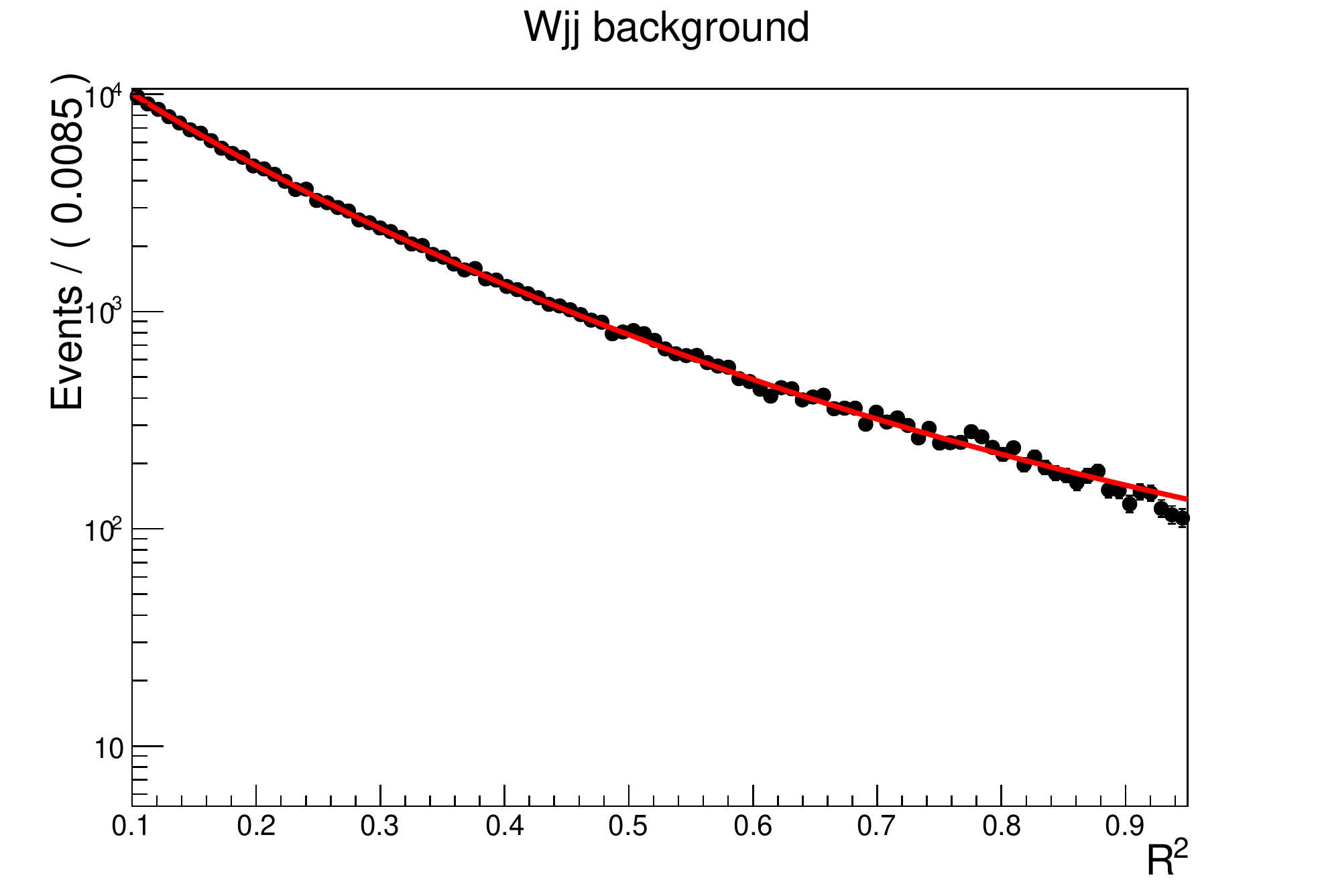} 
\includegraphics[width=0.24\textwidth]{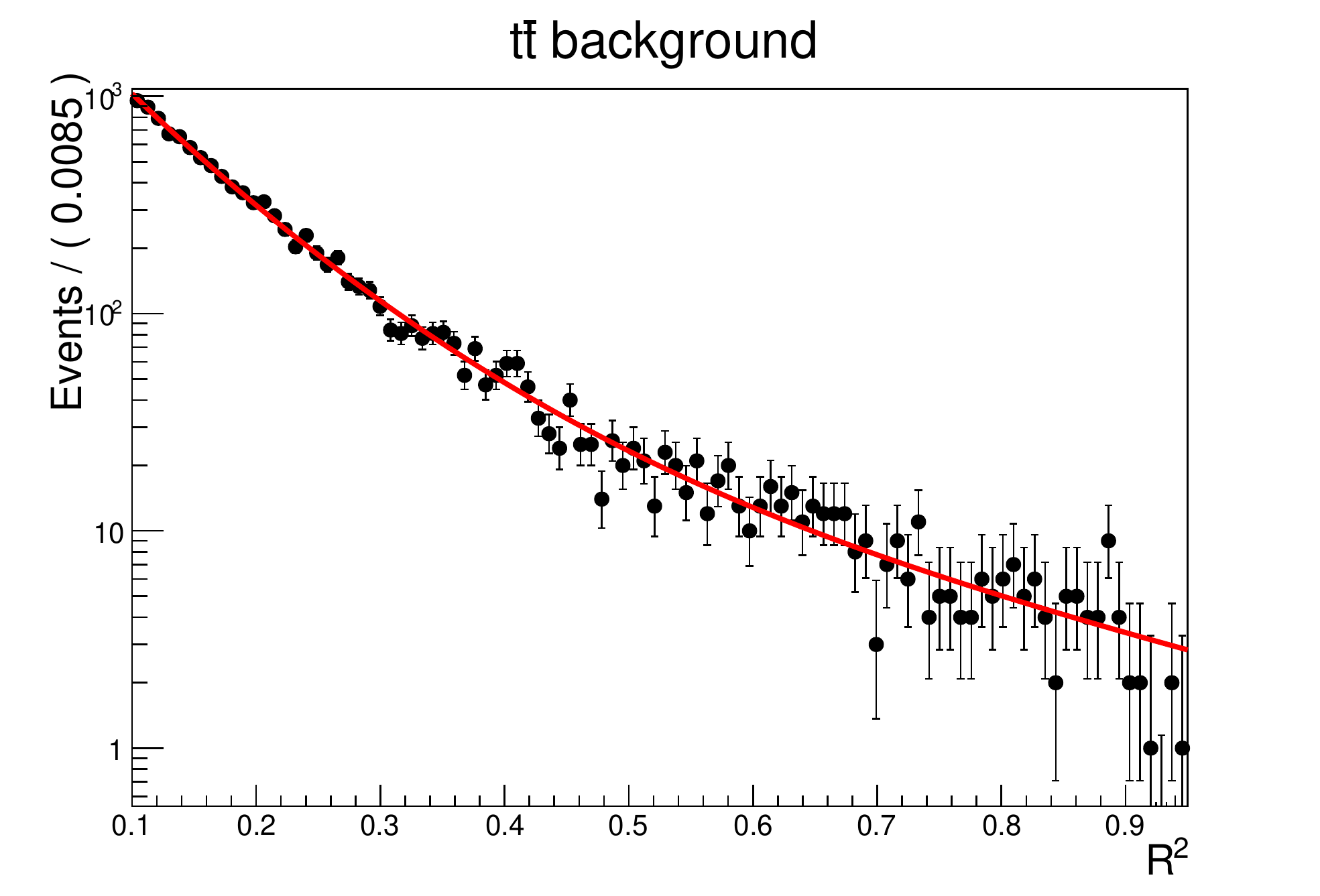}
\includegraphics[width=0.24\textwidth]{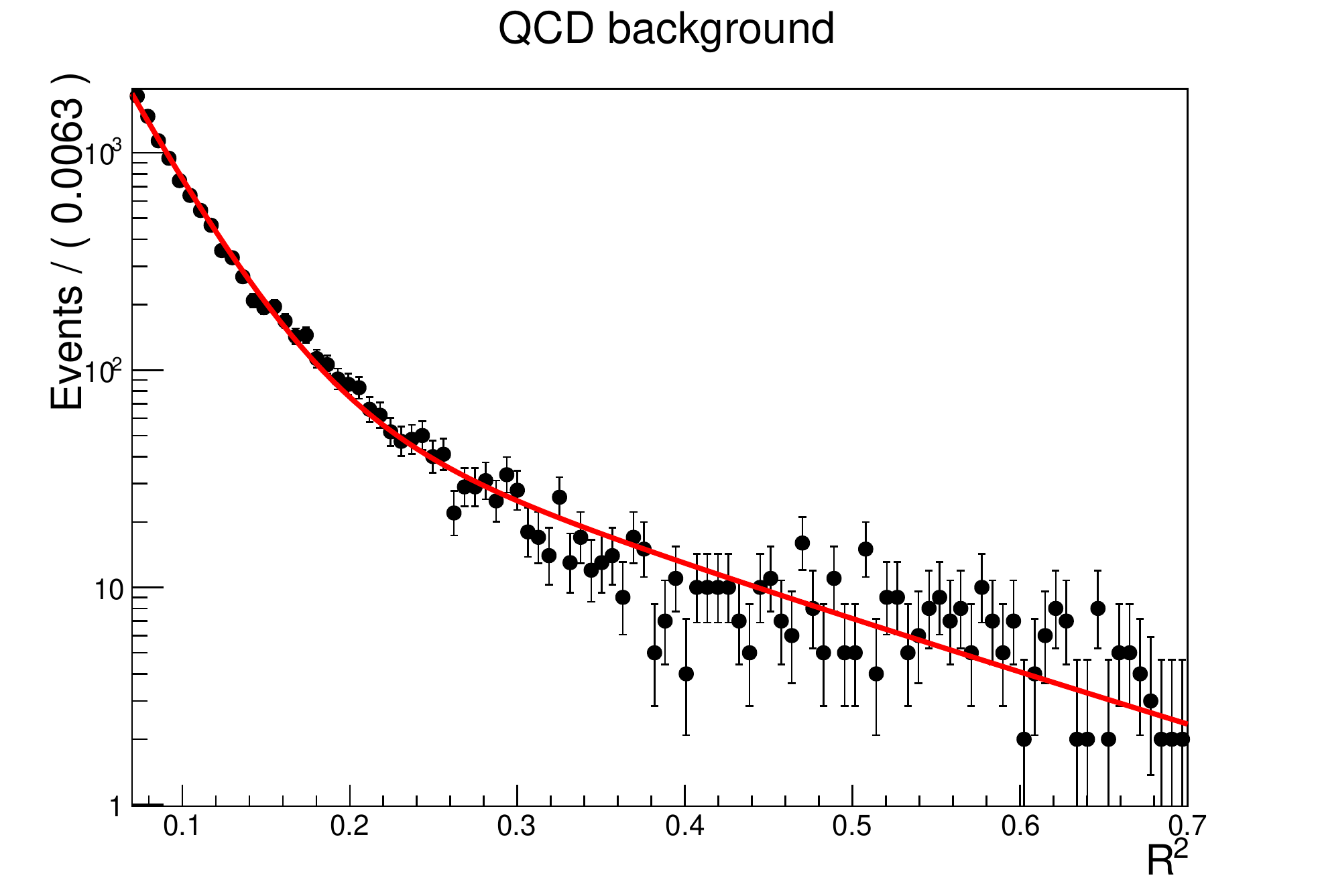} 
\caption{\label{fits} The projected 2-dimensional fit function on top of Monte Carlo data for all backgrounds. Upper panels: project to the $M_R$ integrating $R^2$; lower panels: project to $R^2$ integrating $M_R$. }
\end{figure}

We first apply the preselection cuts (i)-(iii) introduced above to all background events. 
Different fitted regions on the $M_R$-$R^2$ plane are defined for different backgrounds, based on the criteria that the shapes of the variables are smooth and the events are sufficient.
The fitted regions are listed in the second row of Tab.~\ref{tab:bgfit}.
At last, we fit the probability distribution function Eq.~\eqref{pdf} for each background within the fitted region, by using the RooFit toolkit~\cite{Verkerke:2003ir}. 
The fitted parameters are given in Tab.~\ref{tab:bgfit} as well. The estimated distance to minimum (EDM) defined as $2 \cdot \text{EDM} = g^T V g$ where $g$ is gradient and $V$ is covariance matrix shows the convergence of minimization.
In Fig.~\ref{fits}, we plot the projected 2-dimensional fit function on top of MC data (with only statistical uncertainty) for all backgrounds. We can observe that the distributions of backgrounds in $M_R$ and $R^2$ match the probability function quite well.

\begin{table}[htb]
\center
  \begin{tabular}{|c|c|c|c|c|c|c|c|c|}
  \hline
 \multicolumn{2}{|c|}{} & $Zjj$ & $Wjj$ & $t\bar{t}$ & QCD  \\ \hline\hline
 \multirow{2}{*}{Fit region}  & $M_R$ (GeV) & $>200$  & $>200$ & $>300$ & $>150$ \\
  & $R^2$ & $>0.1$ & $>0.1$ & $> 0.1$ & $>0.07$ \\ \hline
 \multirow{7}{*}{Fit parameters}  & $k$ & $2.1317 \times 10^{-2}$ & $1.1756 \times 10^{-2}$ & $4.1768 \times 10^{-2}$ & $1.5270 \times 10^{-1}$ \\
 & $x_0$ & $5.0621 \times 10^{+1}$ & $9.7776 \times 10^{+1}$ & $8.4142 \times 10^{+1}$ & $-2.6242 \times 10^{+1}$ \\
 & $y_0$ & $-2.0950\times 10^{-1}$ & $-2.3448 \times 10^{-1}$ & $-1.1839 \times 10^{-1}$ & $-1.1018 \times 10^{-1}$ \\
 & $f$ & $8.2401 \times 10^{-1}$ & $2.0666 \times 10^{-1}$ & $7.7605 \times 10^{-1}$ & $7.7739 \times 10^{-1}$ \\
 & $k'_0$ & $7.9511\times 10^{-3}$ & $3.1979 \times 10^{-2}$ & $2.0528 \times 10^{-2}$ & $2.9341\times 10^{-2}$  \\
 & $x'_0$ & $1.9719 \times 10^{+2}$ & $4.4126 \times 10^{+1}$  & $1.8022 \times 10^{+2}$ & $-2.5211$ \\
  & $y'_0$  & $-2.8640 \times 10^{-1}$ & $-1.6463 \times 10^{-1}$ & $-9.7141 \times 10^{-2}$ & $-2.9746 \times 10^{-1}$ \\ \hline
   \multicolumn{2}{|c|}{EDM} & $5.7 \times 10^{-4}$ & $7.3 \times 10^{-5}$ & 0.0017 & $3.8 \times 10^{-4}$ \\ \hline
  \multicolumn{2}{|c|}{$\hat{\sigma}_B$ (fb)} & 8.44 & 8.28 & 194.02 & $1.95\times 10^{5}$
  \\ \hline
  \end{tabular}
\caption{\label{tab:bgfit}  Fit region and fitted parameters for each background. The last row gives the corresponding production cross sections of backgrounds to the unit area of the fitted functions. }
 \end{table}

The fitted functions have not been normalized yet, and it can be done as follows. 
For a certain background with the production cross section of $\sigma^0_B$, the total number of simulated events $N_B^0$, the number of events in fitting $N'$ and the area of the fitted function in the fit region $S'$, its production cross section per unit area of the fitted function $\hat{\sigma}_B$ can be calculated as 
\begin{align}
\hat{\sigma}_B = \sigma^0_B \frac{N'}{N^0_B S'},
\end{align}
whose values are given in the last row of Tab.~\ref{tab:bgfit}. 
In a given region $S$ on the $M_R -R^2$ plane, the normalized background cross section $\sigma_B$ is
\begin{align}
\sigma_B = \hat{\sigma}_B  \iint _S P(R^2, M_R) \cdot d M_R ~dR^2~. \label{bbana}
\end{align}

Having obtained the full information of backgrounds in their analytic forms, we can optimize the cuts at each benchmark point to gain the highest sensitivity. 
The optimization is based on the simplified model with $Y_b$=1. 
Their production cross sections are given in Tab.~\ref{tab:xsec14bb}.  

\begin{table}[htb]
\center
  \begin{tabular}{|c|c|c|c|c|c|c|c|c|}
  \hline
  $m_A$(GeV) & 100 & 200 & 300 & 500 & 700 & 1000 & 1500 & 2000 \\ \hline \hline
$ \sigma^{\text{NW}}_{bb}$ (pb) & 119.2 & 21.55 & 6.07 & 0.90 & 0.21 & 0.036 & 0.0036 & $5.4 \times 10^{-4}$  \\
$ \sigma^{\text{WW}}_{bb}$ (pb) & 43.74 & 7.91 & 2.28 & 0.36 & 0.091 & 0.018 & 0.0026 & $6.2 \times 10^{-4}$  \\ \hline
  \end{tabular}
\caption{\label{tab:xsec14bb} 
Benchmark points production cross sections in the NW and WW scenarios at 14 TeV LHC. The two $b$ quarks at parton level are required to have $p_T(b)>20$ GeV, $|\eta(b)|<2.5$ and $\Delta R(b_1,b_2)>0.4$. 
We have set $Y_b$=1 and $m_\chi=1$ GeV. }
 \end{table}

For each benchmark point, after the preselection cuts (i)-(iv), we further require the razor variables to have $M_R> M'_R$ and $R^2 > R'^2$, where $M'_R$ and $R'^2$ are scanned in the selected range of $[300,2100]$ GeV and $[0.1,0.9]$ with step sizes of 200 GeV and 0.1, respectively. 
The signal production cross section for a certain benchmark point after the final selection is $\sigma_S = \sigma_{bb} \times \frac{N_S}{N_S^0}$, where $\sigma_{bb}$ is given in Tab.~\ref{tab:xsec14bb}, $N_S^0$ is the total number of simulated signal events and $N_S$ is the number of signal events after the final selection. 
The corresponding background cross section in that region can be calculated directly via Eq.~(\ref{bbana}). 
The values of $M'_R$ and $R'^2$ are chosen such that $\sigma_S/\sqrt{\sigma_B}$ is maximized. Meanwhile, the ratio of the signal to background cross section $\sigma_S/\sigma_B$ in the selected region is required to be greater than 1\% for the sake of tolerating a relatively large systematic uncertainty.~\footnote{In the next section, we also show the results requiring $\sigma_S/\sigma_B > 5\%$ to show its influence to the search sensitivity.} 
Moreover, $\sigma_S$ should be larger than $10^{-2}$ fb to control the statistical uncertainty.

The $M'_R$ and $R'^2$ cuts for two benchmark points with mediator masses of 100 GeV and 500 GeV in the NW scenario and WW scenario are given in Tab.~\ref{tab:eff14bb}, where the corresponding signal selection efficiency $\epsilon_S$ and the background cross sections after the selection are also listed. 
For the benchmark point with $m_A=100$ GeV, a mild cut of $M_R \gtrsim 300$ GeV is applied, leaving the QCD multi-jets process as the dominant background.
Such a loose cut already helps to suppress the background cross sections to ${\cal O}(1)\,{\rm pb}$. 
While for the other benchmark point, a much harder cut of $M_R \gtrsim 700$ GeV can be applied, and the remaining background is dominated by the $t\bar{t}$ process, whose energy scale is much higher than the QCD multi-jets process. 
This strong cut reduces the background cross section to ${\cal O}(1)$ fb level. 
In the last column of Tab.~\ref{tab:eff14bb}, the signal significances with the integrated luminosity of 3000 fb$^{-1}$ are given. 
Hopefully, the HL-LHC is able to probe the mediator mass up to around 500 GeV (in the WW scenario) with high signal significance as long as $Y_b \sim 1$.

\begin{table}[htb]
\center
  \begin{tabular}{|c|c|c|c|c|c|c|c|c|}
  \hline
   &        $M'_R$(GeV) &  $R'^2$   & $\epsilon_S$  & $\sigma_{Zjj}$(fb)  &  $\sigma_{Wjj}$(fb)  & $\sigma_{\text{QCD}}$(fb)  & $\sigma_{t\bar{t}}$(fb) & $S/\sqrt{S+B}$ \\ \hline \hline
 100NW & $300$  & $0.1$ & $5.03 \times 10^{-3}$ & 177.5 & 53.5 & 9975.4 & 1407.0 & 297 \\
 500NW & $700$  & $0.1$ & $4.39 \times 10^{-3}$ & 18.4 & 4.4 & 45.0 & 91.8 & 16.9 \\
 100WW & $300$  & $0.1$ & $4.28 \times 10^{-3}$ & 177.5 & 53.5 & 9975.4 & 1407.0  & 94.4 \\
 500WW & $900$ & $0.1$ & $1.64 \times 10^{-3}$ & 7.5 & 1.6 & 3.5 & 31.1 & 4.9 \\ \hline
  \end{tabular}
\caption{\label{tab:eff14bb} Tables for cuts on the razor variables (2nd and 3nd columns); cross sections of the backgrounds after the final selection (5nd to 8nd column);  signal significance at 14 TeV LHC with integrated luminosity 3000 fb$^{-1}$ (last column). We show two benchmark points with $m_A=100/500$ GeV, in the NW and WW scenarios.}
 \end{table}

\section{Exclusion limit on $Y_b$ at 14 TeV LHC}
\label{sec:results}

\begin{figure}[htb]
\includegraphics[width=0.48\textwidth]{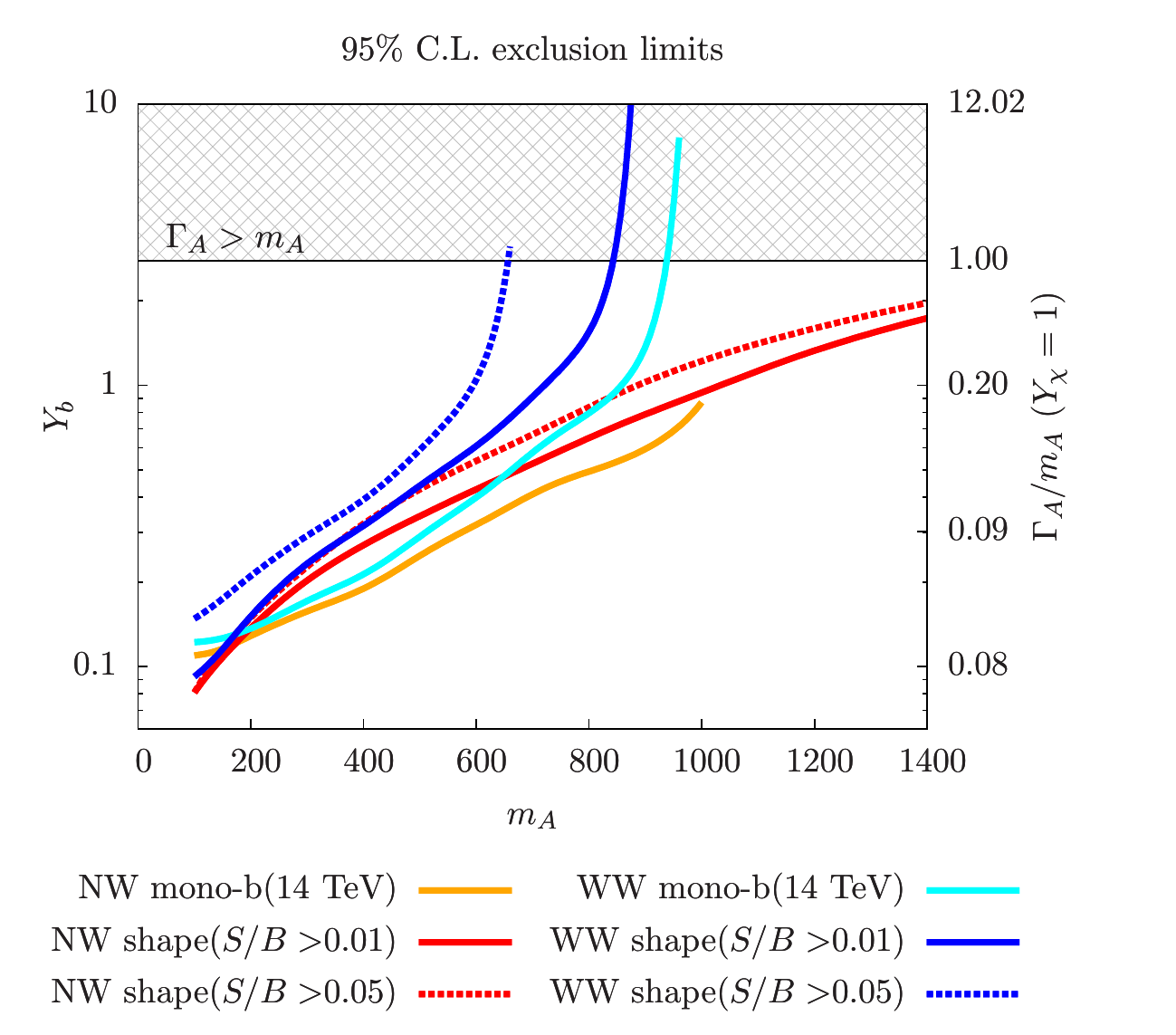}
\includegraphics[width=0.48\textwidth]{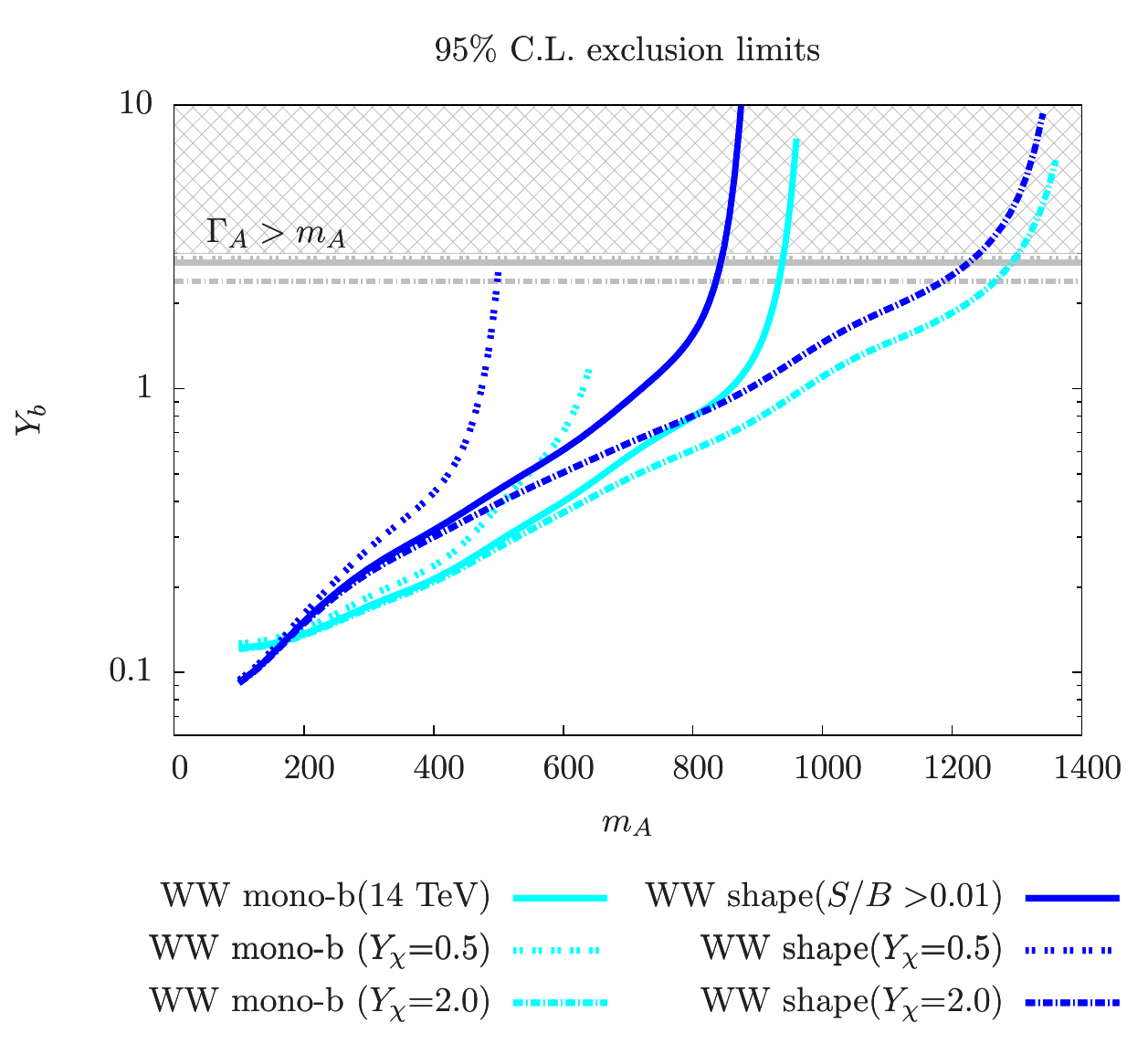}
\caption{\label{fig:exclusion} The 95\% CL exclusion limits on $Y_b$ in terms of the mono-$b$ analysis  and shape analysis at 14 TeV 3000 fb$^{-1}$ HL-LHC;  the latter analysis is demonstrated for two cases, respectively having the signal to background ratio greater than 1\% and 5\%. The left panel ($Y_\chi=1$) is for comparing the NW and WW scenarios while the right is for showing the influence of varying $Y_\chi$ in the WW scenario. The region of $\Gamma_A > m_A$ is marked with grid in both panels. In the left panel, we also show the $\Gamma_A /m_A$ as a second Y-axis because it is monotonically determined by $Y_b$ in the region $m_A \gg m_b, m_\chi$.  In the right panel, different lower bounds of grid correspond to the different $Y_\chi$ with the same line type. }
\end{figure}

In Sec.~\ref{sec:monob}, we gave the 95\% CL upper limit on the production cross section $\sigma^{8/14}_S$ (with arbitrary integrated luminosity) by using the mono-$b$ analysis, while in Sec.~\ref{sec:razor} we showed the signal significances $S/\sqrt{S+B}$  (with ${\cal L}= 3000\,{\rm fb}^{-1}$) by using the shape analysis with respect to 2$b$-jets. 
They can be converted into bounds on $Y_b$ after fixing $Y_\chi$  and $m_A$. For the mono-$b$ case, one obtains
\begin{align}
Y^{\text{NW}}_b(\text{mono-$b$}) &=  \left(\frac{\sigma^{8/14}_S}{\sigma^{\text{NW}}_{p_T(b_1)>50~\text{GeV}}}\right)^{1/2}, 
\cr
Y^{\text{WW}}_b(\text{mono-$b$}) &=\left(\f{5}{2} \frac{\sigma^{\text{WW}}_{p_T(b_1)>50~\text{GeV}}}{\sigma^{8/14}_S} -\f{3}{2} Y_{\chi}^{-2}\right)^{-1/2}. 
\end{align}
In the WW scenario, ${\sigma^{\text{WW}}_{p_T(b_1)>50~\text{GeV}}}/{\sigma^{8/14}_S}\gtrsim 0.6 (1/Y_\chi)^2$ is needed since $\sigma^{\text{WW}}(Y_b)$ is bounded from above when increasing $Y_b$. As for the 2$b$-jets case, one gets the 95\% CL (corresponding to 2-$\sigma$ level signal significance) upper limit on $Y_b$ as
\begin{align}
Y^{\text{NW}}_b(\text{shape}) &= \sqrt{\frac{2}{S/\sqrt{S+B}}}, \quad
Y^{\text{WW}}_b(\text{shape}) = \left( {\f{5}{2} \frac{S}{2\sqrt{S+B}} -\f{3}{2} Y_{\chi}^{-2}}\right)^{-1/2}.
\end{align}
The results are displayed in Fig.~\ref{fig:exclusion}, and we find that in the NW scenario our results are well consistent with those of Ref.~\cite{Berlin:2015wwa}.

From the figures we get a few observations. 
First, in interpreting the searches, the NW scenario yield much more stringent bounds on $Y_b$ than the realistic WW scenario. 
For heavy inputs of $m_A$, the NW scenario may turn out to be far from reliable, and the reasons have been explained before. 
Second, in the relatively heavier mediator region (far above 200 GeV), the mono-$b$ search would yield significantly stronger bound on $Y_b$ than the 2$b$ search.~\footnote{Under the condition that the signal to background ratio to be great than a few percent. 
Nevertheless, in the region of $m_A\lesssim 200$ GeV, the 2$b$ search (equipped with shape analysis) is competitive and even better than the mono-$b$ search. 
One can understand these from the relative cut efficiencies in Tab.~\ref{xsec}. 
For smaller $m_A$, the cut of large MET is more stringent than two $b$-tagged jets.
However, the situation rapidly reverses as $m_A$ increases. 
Third, the effects from changing $Y_{\chi}$ become significant only in the region of heavy $m_A$.
As expected, a larger $Y_\chi$ will lead to a stronger bound on $Y_b$. } 
Third, as expected, the parameter space of interest has a rather large $Y_b$ in particular. 
Therefore, flavor physics may raise the question of if our search is of real interest. 
For instance, the most stringent constraint from $B\ra X_s+\gamma$ imposes a lower bound on the mass of the charged Higgs boson in the type-II 2HDM of $m_{H^+}\gtrsim 485$ GeV, which is almost independent on $Y_b$~\cite{Misiak:2015xwa}. 
Even so, there is still a large parameter space remaining for our search. Moreover, that kind of bound does not apply to models beyond the minimal type-II 2HDM where the charged Higgs boson mass is not tied to the neutral mediator mass.

\section{Conclusion}
\label{sec:conclusion}

In this work, we have analyzed the LHC signatures for the Type-II 2HDM-like Higgs Portal DM model. 
The model has sizable production rate of the $gg\to b\bar b \Phi_1(\to \chi \chi)$ process in the large $\tan\beta$ region. 
Thus can be searched for in final states containing either one energetic $b$-tagged jet plus MET or two $b$-tagged jets plus MET, depending on the size of the transverse momentum of the second $b$-jet.

At the particle level, for $m_{H/A} \gtrsim 125$ GeV,  there are more than 37\% of signal events that have at least one b-quark with $p_T(b) >20$ GeV and $|\eta(b)|<2.5$. The efficiency drops to 2.5\% when large MET ($p_T(H/A)>100$ GeV) is required.  The mono-$b$ signature has been searched at the LHC based on the effective operator $\mathcal{O}_b$. 
We recast the experimental analysis in our $\Phi_1$-like portal DM model. The exclusion bound is extrapolated to 14 TeV LHC with integrated luminosity of 3000 fb$^{-1}$.  We find that in the light mediator region, for a wide range of $Y_\chi$, models with $Y_b$ as small as $\sim 0.1$ can be probed/excluded at the HL-LHC. 

The efficiency for two b-quarks signal is around 5\% for $m_{H/A} =125$ GeV, and can be increased to $\sim 17\%$ when $m_{H/A} \sim 1$ TeV. The information of the additional $b$-jet in the $2b$+MET signature can help to suppress the backgrounds without requiring large MET, thus will improve the search sensitivity in the light mediator region comparing to the mono-$b$ signature. 
We adopt a search for final state containing exactly two $b$-tagged jets using the razor variables. 
The distributions of which for SM backgrounds can be simply modeled by smooth functions, so that heavy use Monte Carlo simulation can be avoided. 
By studying the shapes of the razor variables, we find the $2b$+MET search has comparable sensitivity with the mono-$b$ search when requiring the signal to background ratio to be greater than a few percent. Especially, for $m_{H/A} \sim 125$ GeV, even $Y_b$ smaller than 0.1 can be reached by the HL-LHC with very mild dependence on $Y_\chi$.


\section*{ACKNOWLEDGMENTS}

We would like to thank Xiaogang He for very useful discussions and communication. 
N.C. is partially supported by the National Science Foundation of China (under Grant No. 11575176), the Fundamental Research Funds for the Central Universities (under Grant No. WK2030040069).
We would like to thank the Kavli Institute for Theoretical Physics China at the Chinese Academy of Sciences for their hospitalities when part of this work was prepared.


\end{document}